\documentclass[ALICE,manyauthors]{cernphprep}
\pdfoutput=1
\usepackage[comma,square,numbers,sort&compress]{natbib}
\usepackage{hyperref}
\usepackage{lineno}
\usepackage[T1]{fontenc} 
\usepackage{setspace}
\usepackage{capt-of}
\usepackage{epstopdf}
\usepackage{color}
\usepackage{multirow}
\usepackage{tabu}

\newcommand{\ITS}{\rm{ITS}}
\newcommand{\SPD}{\rm{SPD}}
\newcommand{\SDD}{\rm{SDD}}
\newcommand{\SSD}{\rm{SSD}}
\newcommand{\TPC}{\rm{TPC}}

\newcommand{\TOF}{\rm{TOF}}
\newcommand{\VZERO}{\rm{V0}}

\newcommand{\VZEROA}{\rm{V0A}}
\newcommand{\VZEROC}{\rm{V0C}}

\newcommand{\pT}{$p_{\mathrm{T}}$}

\newcommand{\pTnq}{$p_{\mathrm{T}}/n_{q}$}
\newcommand{\pTmT}{$(m_{\mathrm{T}}-m_{0})/n_{q}$}

\newcommand{\rawvtwo}{$v_{2}$}
\newcommand{\rawvthree}{$v_{3}$}
\newcommand{\rawvfour}{$v_{4}$}

\newcommand{\rawvn}{$v_{n}$}

\newcommand{\nq}{$n_{q}$}

\newcommand{\vtwo}{$v_{2}^{\mathrm{sub}}$}
\newcommand{\vthree}{$v_{3}^{\mathrm{sub}}$}
\newcommand{\vfour}{$v_{4}^{\mathrm{sub}}$}
\newcommand{\vfive}{$v_{5}^{\mathrm{sub}}$}
\newcommand{\vn}{$v_{n}^{\mathrm{sub}}$}

\newcommand{\vtwoAA}{$v_{2}^{\mathrm{AA}}$}
\newcommand{\vthreeAA}{$v_{3}^{\mathrm{AA}}$}
\newcommand{\vfourAA}{$v_{4}^{\mathrm{AA}}$}
\newcommand{\vfiveAA}{$v_{5}^{\mathrm{AA}}$}
\newcommand{\vnAA}{$v_{n}^{\mathrm{AA}}$}

\newcommand{\vtwonq}{$v_{2}^{\mathrm{sub}}/n_{q}$}
\newcommand{\vthreenq}{$v_{3}^{\mathrm{sub}}/n_{q}$}
\newcommand{\vfournq}{$v_{4}^{\mathrm{sub}}/n_{q}$}
\newcommand{\vfivenq}{$v_{5}^{\mathrm{sub}}/n_{q}$}

\newcommand{\deltatwo}{$\delta_{2}^{\textrm{AA,pp}}$}
\newcommand{\deltathree}{$\delta_{3}^{\textrm{AA,pp}}$}
\newcommand{\deltafour}{$\delta_{4}^{\textrm{AA,pp}}$}
\newcommand{\deltafive}{$\delta_{5}^{\textrm{AA,pp}}$}
\newcommand{\deltan}{$\delta_{n}^{\textrm{AA,pp}}$}

\newcommand{\pion}{$\pi^{\pm}$}
\newcommand{\kaon}{$\mathrm{K}^{\pm}$}
\newcommand{\proton}{p+$\overline{\mathrm{p}}$}

\newcommand{\GeV}{$\mathrm{GeV}/c$}
\newcommand{\sNN}{$\sqrt{s_{\mathrm{NN}}}=2.76$ TeV}

\newcommand{\uQpp}{$\langle{M}\rangle^{\textrm{pp}}\langle{\langle{\vec{u}_{n}\cdot\frac{\vec{Q}^{*}_{n}}{M}}\rangle}\rangle^{\rm{pp}}$}

\newcolumntype{?}{!{\vrule width 2pt}}
\newcolumntype{@}{!{\vrule width 1.5pt}}

\begin{document}%
\maketitle

\begin{titlepage}
\PHyear{2016}
\PHnumber{159}      
\PHdate{15 June}  
%

\title{Higher harmonic flow coefficients of identified hadrons in Pb--Pb collisions at $\mathbf{\sqrt{s_\mathrm{{\mathbf{NN}}}}}$~=~2.76~TeV}
\ShortTitle{Higher harmonic flow at the LHC}   

\Collaboration{ALICE Collaboration\thanks{See Appendix~\ref{app:collab} for the list of collaboration members}}
\ShortAuthor{ALICE Collaboration} 

\begin{abstract}
The elliptic, triangular, quadrangular and pentagonal anisotropic flow coefficients for $\pi^{\pm}$, $\mathrm{K}^{\pm}$ and p+$\overline{\mathrm{p}}$ in Pb--Pb collisions at $\sqrt{s_\mathrm{{NN}}} = 2.76$~TeV were measured with the ALICE detector at the Large Hadron Collider. The results were obtained with the Scalar Product method, correlating the identified hadrons with reference particles from a different pseudorapidity region. Effects not related to the common event symmetry planes (non-flow) were estimated using correlations in pp collisions and were subtracted from the measurement. The obtained flow coefficients exhibit a clear mass ordering for transverse momentum (\pT) values below $\approx$ 3~GeV/$c$. In the intermediate \pT~region ($3 < p_{\mathrm{T}} < 6$~GeV/$c$), particles group at an approximate level according to the number of constituent quarks, suggesting that coalescence might be the relevant particle production mechanism in this region. The results for $p_{\mathrm{T}} < 3$~GeV/$c$ are described fairly well by a hydrodynamical model (iEBE-VISHNU) that uses initial conditions generated by A Multi-Phase Transport model (AMPT) and describes the expansion of the fireball using a value of 0.08 for the ratio of shear viscosity to entropy density ($\eta/s$), coupled to a hadronic cascade model (UrQMD). Finally, expectations from AMPT alone fail to quantitatively describe the measurements for all harmonics throughout the measured transverse momentum region. However, the comparison to the AMPT model highlights the importance of the late hadronic rescattering stage to the development of the observed mass ordering at low values of $p_{\mathrm{T}}$ and of coalescence as a particle production mechanism for the particle type grouping at intermediate values of $p_{\mathrm{T}}$ for all harmonics.
\end{abstract}
\setcounter{page}{1}
\end{titlepage}
\setcounter{page}{2}
\tableofcontents
\newpage
\setcounter{page}{3}

\section{Introduction}
\label{Sec:Introduction}
Quantum chromodynamics (QCD) calculations on the lattice~\cite{Borsanyi:2010cj,Bhattacharya:2014ara} suggest that at high values of temperature and energy density a transition takes place from ordinary nuclear matter to a state where the constituents, the quarks and the gluons, are deconfined. This state of matter is called the quark-gluon plasma (QGP)~\cite{Shuryak:1984nq,Cleymans:1985wb,Bass:1998vz}. The aim of the heavy-ion program at the Large Hadron Collider (LHC) is to study the QGP properties, such as the equation of state, the speed of sound in the medium, and the value of the ratio of shear viscosity to entropy density ($\eta/s$).

One of the important observables sensitive to the properties of the QGP is the azimuthal distribution of particles emitted in the plane transverse to the beam direction. In non-central collisions between two heavy ions the overlap region is not isotropic. This spatial anisotropy of the overlap region is transformed into an anisotropy in momentum space initially through interactions between partons and at later stages between the produced particles. The resulting anisotropy is usually expressed in terms of a Fourier series in azimuthal angle $\varphi$~\cite{Voloshin:1994mz,Poskanzer:1998yz} according to

\begin{equation}
E\frac{\mathrm{d}^3N}{\mathrm{d}p^3} = \frac{1}{2\pi}\frac{\mathrm{d}^2N}{p_{\mathrm{T}}\mathrm{d}p_{\mathrm{T}}\mathrm{d}\eta} \Big\{1 + 2\sum_{n=1}^{\infty} v_n(p_{\mathrm{T}},\eta) \cos[n(\varphi - \Psi_n)]\Big\},
\label{Eq:Fourier}
\end{equation}

\noindent where $E$, $N$, $p$, $p_{\mathrm{T}}$, $\varphi$ and $\eta$ are the energy, particle yield, total momentum, transverse momentum, azimuthal angle and pseudorapidity of particles, respectively, and $\Psi_n$ is the azimuthal angle of the symmetry plane of the $n^{\mathrm{th}}$-order harmonic~\cite{Bhalerao:2006tp,Alver:2008zza,Alver:2010gr,Alver:2010dn}. The $n^{\mathrm{th}}$-order flow coefficients are denoted as $v_n$ and can be calculated as

\begin{equation}
v_{n} = \langle{\cos[n(\varphi - \Psi_n)]}\rangle,
\label{Eq:vn}
\end{equation}

where the brackets denote an average over all particles in all events. Since the symmetry planes are not accessible experimentally, the flow coefficients are estimated solely from the azimuthal angles of the produced particles. The second Fourier coefficient, $v_2$, measures the elliptic flow, i.e.~the momentum space azimuthal anisotropy of particle emission relative to the second harmonic symmetry plane. The study of $v_2$~at both the Relativistic Heavy Ion Collider (RHIC) and the LHC contributed significantly to the realisation that the produced system can be described as a strongly-coupled quark-gluon plasma (sQGP) with a small value of $\eta/s$, very close to the conjectured lower limit of $1/4\pi$ from AdS/CFT~\cite{Kovtun:2004de}. 

In addition, the overlap region of the colliding nuclei exhibits an irregular shape~\cite{Manly:2005zy,Bhalerao:2006tp,Alver:2008zza,Alver:2010gr,Alver:2010dn}. The irregularities originate from the initial density profile of nucleons participating in the collision, which is not isotropic and differs from one event to the other. This, in turn, causes the symmetry plane of the irregular shape to fluctuate in every event around the reaction plane, defined by the impact parameter vector and the beam axis, and also gives rise to the additional higher harmonic symmetry planes $\Psi_n$. The initial state fluctuations yield higher order flow harmonics such as $v_3$, $v_4$, and $v_5$ that are usually referred to as triangular, quadrangular, and pentagonal flow, respectively. Recent calculations~\cite{Teaney:2010vd,Qin:2010pf} suggest that their transverse momentum dependence is a more sensitive probe than elliptic flow not only of the initial geometry and its fluctuations, but also of $\eta/s$. The first measurements of the $p_{\mathrm{T}}$-differential $v_n$, denoted as $v_n(p_{\mathrm{T}})$, of charged particles at the LHC~\cite{ALICE:2011ab,ATLAS:2012at,Chatrchyan:2013kba} provided a strong testing ground for hydrodynamical calculations that attempt to describe the dynamical evolution of the system created in heavy-ion collisions. 

An additional challenge for hydrodynamical calculations and a constraint on both the initial conditions and $\eta/s$ can be provided by studying the flow coefficients of Eq.~\ref{Eq:vn} as a function of collision centrality and transverse momentum for different particle species. The first results of such studies at RHIC~\cite{Adams:2003am,Abelev:2007qg,Adler:2003kt,Adare:2006ti} and the LHC~\cite{Abelev:2014pua,Adam:2015eta} revealed that an interplay of radial flow (the average velocity of the system's collective radial expansion) and anisotropic flow leads to a characteristic mass dependence of $v_2(p_{\mathrm{T}})$~\cite{Voloshin:1996nv,Huovinen:2001cy,Shen:2011eg} for $p_{\mathrm{T}} < 3$~GeV/$c$. For higher values of transverse momentum up to $p_{\mathrm{T}} \approx 6$~GeV/$c$ these results indicate that the $v_2$ of baryons is larger than that of mesons. This behaviour was explained in a dynamical model where flow develops at the partonic level followed by quark coalescence into hadrons~\cite{Voloshin:2002wa,Molnar:2003ff}. This mechanism leads to the observed hierarchy in the values of $v_2(p_{\mathrm{T}})$, referred to as number of constituent quarks (NCQ) scaling. New results from ALICE~\cite{Abelev:2014pua} and PHENIX~\cite{Adare:2012vq} exhibit deviations from the NCQ scaling at the level of $\pm$20$\%$ for $p_{\mathrm{T}} > 3$~GeV/$c$. In addition, the LHC results showed also that the $v_2$ of the $\phi$-meson at intermediate values of transverse momentum follows the baryon rather than the meson scaling for central Pb--Pb collisions~\cite{Abelev:2014pua}. Recently, the first results of $v_2(p_{\mathrm{T}})$, $v_3(p_{\mathrm{T}})$, and $v_4(p_{\mathrm{T}})$ for $\pi^{\pm}$, $\mathrm{K}^{\pm}$ and p+$\overline{\mathrm{p}}$ for 50$\%$ most central Au--Au collisions at $\sqrt{s_{\mathrm{NN}}} = 200$~GeV were reported~\cite{Adare:2014kci}. The higher harmonic flow coefficients exhibit similar mass and particle-type dependences as $v_2$ up to intermediate values of $p_{\mathrm{T}}$.

In this article, we report the results for the $p_{\mathrm{T}}$-differential elliptic, triangular, quadrangular and pentagonal flow for \pion, \kaon~and \proton~measured in Pb--Pb collisions at the centre of mass energy per nucleon pair $\sqrt{s_\mathrm{{NN}}} = 2.76$~TeV with the ALICE detector~\cite{Aamodt:2008zz,Abelev:2014ffa} at the LHC. The particles are identified using signals from both the Time Projection Chamber (\TPC) and the Time Of Flight (\TOF) detectors, described in Section~\ref{Sec:ExpSetup}, with a procedure that is discussed in Section~\ref{Sec:AnalysisDetails}. The results are obtained with the Scalar Product method described in Section~\ref{Sec:FlowMethods}, and in detail in Refs.~\cite{Abelev:2014pua,Adler:2002pu,Voloshin:2008dg,vanderKolk:2012oca}. In this article, the identified hadron under study and the charged reference particles are obtained from different, non-overlapping pseudorapidity regions. A correction for correlations not related to the common symmetry plane (non-flow), like those arising from jets, resonance decays and quantum statistics correlations, is presented in Section~\ref{Sec:FlowMethods}. This procedure relies on measuring the corresponding correlations in pp collisions and subtracting them from the $v_n$ coefficients measured in Pb--Pb collisions to form the reported $v_n^{\textrm{sub}}(p_{\mathrm{T}})$, where the superscript $\textrm{`sub'}$ is used to stress the subtraction procedure. The systematic uncertainties of the measurements are described in Section~\ref{Sec:Systematics}. All harmonics were measured separately for particles and anti-particles and were found to be compatible within the statistical uncertainties. Therefore, the $v_n^{\textrm{sub}}(p_{\mathrm{T}})$ for the average of the results for the opposite charges is reported. The results are reported in Section~\ref{Sec:Results} for the 0--50$\%$ centrality range of Pb--Pb collisions. Finally, results are also reported separately for ultra-central events, i.e.~the 0--1$\%$ centrality range, where the role of the collision geometry is reduced and one expects that $v_n^{\textrm{sub}}(p_{\mathrm{T}})$ is mainly driven by the initial state  fluctuations.

\section{Experimental setup}
\label{Sec:ExpSetup}
ALICE~\cite{Aamodt:2008zz,Abelev:2014ffa} is one of the four large experiments at the LHC, particularly designed to cope with the large charged-particle densities present in central Pb--Pb collisions~\cite{Aamodt:2010pb}. By convention, the beam direction defines the $z$-axis, the $x$-axis is horizontal and points towards the centre of the LHC, and the $y$-axis is vertical and points upwards. The apparatus consists of a set of detectors located in the central barrel, positioned inside a solenoidal magnet which generates a $0.5$~T field parallel to the beam direction, and a set of forward detectors. 

The Inner Tracking System (\ITS)~\cite{Aamodt:2008zz} and the \TPC~\cite{Alme:2010ke} are the main tracking detectors of the central barrel. The \ITS~consists of six layers of silicon detectors employing three different technologies. The two innermost layers, positioned at $r = 3.9$~cm and 7.6~cm,  are Silicon Pixel Detectors (\SPD), followed by two layers of Silicon Drift Detectors (\SDD) ($r = 15$~cm and 23.9~cm). Finally, the two outermost layers are double-sided Silicon Strip Detectors (\SSD) at $r = 38$~cm and 43~cm. The \TPC~surrounds the \ITS~and provides full azimuthal coverage in the pseudorapidity range $|\eta| < 0.9$. 

Charged pions, kaons and protons were identified using the information from the \TPC~and the \TOF~detectors~\cite{Aamodt:2008zz}. The \TPC~allows for a simultaneous measurement of the momentum of a particle and its specific energy loss $\langle \mathrm{d}E/\mathrm{d}x \rangle$ in the gas. The detector provides a separation by at least 2 standard deviations for the hadron species at $p_{\mathrm{T}} < 0.7$~GeV/$c$ and the possibility to identify particles in the relativistic rise region of $\mathrm{d}E/\mathrm{d}x$ (i.e.~$2 < p_{\rm{T}} < 20$~GeV/$c$)~\cite{Abelev:2014ffa}. The $\mathrm{d}E/\mathrm{d}x$ resolution for the 5$\%$ most central Pb--Pb collisions is 6.5$\%$ and improves for more peripheral collisions. The \TOF~detector is placed around the \TPC~and provides a $3\sigma$ separation between $\pi$--K and K--$\rm{p}$ up to $p_{\mathrm{T}} = $ 2.5~GeV/$c$ and $p_{\mathrm{T}} = 4$~GeV/$c$, respectively~\cite{Abelev:2014ffa}. This is done by measuring the flight time of particles from the collision point with a resolution of about $80$~ps. The start time for the \TOF~measurement is provided by the T0 detectors, two arrays of Cherenkov counters positioned at opposite sides of the interaction points covering $4.6 < \eta < 4.9$ (T0A) and $-3.3 < \eta < -3.0$ (T0C). The start time is also determined using a combinatorial algorithm that compares the timestamps of particle hits measured by the TOF to the expected times of the tracks, assuming a common event time $t_{ev}$ \cite{Abelev:2014ffa}. Both methods of estimating the start time are fully efficient for the 50$\%$ most central Pb--Pb collisions.

A set of forward detectors, the \VZERO~scintillator arrays~\cite{Abbas:2013taa}, were used in the trigger logic and for the determination of the collision centrality, discussed in the next section. The \VZERO~consists of two systems, the \VZEROA~and the \VZEROC, that are positioned on each side of the interaction point and cover the pseudorapidity ranges of $2.8 < \eta < 5.1$ and $-3.7 < \eta < -1.7$, respectively. 

For more details on the ALICE experimental setup and the performance of the detectors, see Refs.~\cite{Aamodt:2008zz,Abelev:2014ffa}.

\section{Event sample, track selection and particle identification}
\label{Sec:AnalysisDetails}

\subsection{Trigger selection and data sample}
\label{SubSec:Events}
   
The analysis is performed on data from pp and Pb--Pb collisions at $\sqrt{s_{\mathrm{NN}}} = 2.76$~TeV collected with the ALICE detector in 2011. The minimum bias trigger in pp collisions required at least one hit in either of the \VZERO~detectors or the \SPD. In Pb--Pb collisions, minimum bias events were triggered by the coincidence between signals from the two sides of the \VZERO~detector. In addition, in Pb--Pb collisions, an online selection based on the \VZERO~detectors was used to increase the number of central (i.e.~0--10$\%$ centrality range) and semi-central (i.e.~10--50$\%$ centrality range) events. An offline event selection, exploiting the signal arrival time in \VZEROA~and \mbox{\VZEROC}, measured with a 1~ns resolution, was used to discriminate background (e.g.~beam--gas) from collision events. This led to a reduction of background events in the analysed samples to a negligible fraction ($< 0.1 \%$)~\cite{Abelev:2014ffa}. All events selected for the analysis had a reconstructed primary vertex position along the beam axis ($z_{\textrm{vtx}}$) within 10~cm from the nominal interaction point. Finally, events with multiple reconstructed vertices were rejected, leading to a negligible amount of pile--up events for all systems~\cite{Abelev:2014ffa}. After all the selection criteria, a filtered data sample of approximately $25~\times$ $10^6$ Pb--Pb and $20~\times$ $10^6$ pp events were analysed to produce the results presented in this article. 

Events were classified according to fractions of the inelastic cross section and correspond to the 50$\%$ most central Pb--Pb collisions. The 0--1$\%$ interval represents the most central interactions (i.e.~smallest impact parameter) and will be referred to as ultra-central collisions in the following. On the other hand, the 40--50$\%$ interval corresponds to the most peripheral (i.e.~largest impact parameter) collisions in the analysed sample, imposed by the usage of the semi-central trigger for the collected sample in 2011. The centrality of the collision was estimated using the distribution of signal amplitudes from the \VZERO~detectors. The systematic uncertainty due to the centrality estimation is determined using the charged particle multiplicity distribution of \TPC~tracks and the number of \SPD~clusters, and will be discussed in Section~\ref{Sec:Systematics}. Details about the centrality determination can be found in Ref.~\cite{Abelev:2013qoq}.

\subsection{Track selection}
\label{SubSec:Selection}
In this analysis, tracks are reconstructed using the information from the \TPC~and the \ITS~detectors. The tracking algorithm, based on the Kalman filter~\cite{Billoir:1983mz,Billoir:1985nq}, starts from a collection of space points (referred to as clusters) inside the \TPC, and provides the quality of the fit by calculating its $\chi^2$ value. Each space point is reconstructed at one of the TPC padrows, where the deposited ionisation energy is also measured. The specific ionisation energy loss $\langle \mathrm{d}E/\mathrm{d}x \rangle$ is estimated using a truncated mean, excluding the 40\% highest-charge clusters associated to the track. The obtained $\langle \mathrm{d}E/\mathrm{d}x \rangle$ has a resolution, which we later refer to as $\sigma_{\mathrm{TPC}}$. The tracks are propagated to the outer layer of the \ITS, and the tracking algorithm attempts to identify space points in each one of the consecutive layers, reaching the innermost ones (i.e.~\SPD). The track parameters are then updated using the combined information from both the \TPC~and the \ITS~detectors. If the algorithm is unable to match the track reconstructed in the \TPC~with associated \ITS~clusters (e.g.~due to inefficiencies caused by dead channels in some of the \ITS~layers), the track parameters calculated from the \TPC~tracking algorithm are used instead. This tracking mode will be referred to as hybrid tracking in the rest of the text, and is used as the default in this analysis since it also provides uniform $\varphi$ distribution. 

Primary charged pions, kaons and (anti-)protons were required to have at least 70 reconstructed space points out of the maximum of 159 in the \TPC. The average $\chi^2$ of the track fit per TPC space point per degree of freedom (see~\cite{Abelev:2014ffa} for details) was required to be below $2$. These selections reduce the contribution from short tracks, which are unlikely to originate from the primary vertex. To further reduce the contamination by secondary tracks from weak decays or from the interaction with the material, only particles within a maximum distance of closest approach (DCA) between the tracks and the primary vertex in both the transverse plane ($\mathrm{DCA}_{xy} < 2.4$~cm) and the longitudinal direction ($\mathrm{DCA}_{z} < 3.2$~cm) were analysed. Moreover, the tracks were required to have at least two associated \ITS~clusters in addition to having a hit in either of the two \SPD~layers. This selection leads to an efficiency of about $80\%$ for primary tracks at $p_\mathrm{T} > 0.6$~GeV/$c$ and a contamination from secondaries of about $5\%$ at $p_{\rm{T}} = 1$~GeV/$c$~\cite{Abelev:2013vea}. These values depend on particle species and transverse momentum~\cite{Abelev:2013vea}. 

The systematic uncertainty due to the track reconstruction mode was estimated using two additional tracking modes, one relying on the so-called standalone \TPC~tracking with the same parameters described before, and a second that relies on the combination of the \TPC~and the \ITS~detectors (i.e.~global tracking) with tighter selection criteria. In the latter case, the maximum value of DCA was 0.3~cm in both the transverse plane and the longitudinal direction, thus further reducing the amount of secondary particles in the track sample.

The results are reported for all identified hadrons in $|\eta| < 0.8$ and for the transverse momentum range $0.3 <$ \pT~$< 6.0$~\GeV~for \pion~and $0.3 < $ \pT~$< 4.0$~\GeV~for \kaon. Finally, since the contamination from secondary protons created through the interaction of particles with the detector material can reach values larger than 5$\%$ for $p_{\rm{T}} < 1$~GeV/$c$, only $\overline{\mathrm{p}}$ were considered for $0.4 < p_{\rm{T}} < 1$~GeV/$c$, while for higher values (i.e.~$1 < p_{\rm{T}} < 6$~GeV/$c$) a combined measurement of p and $\overline{\mathrm{p}}$ is reported.

\subsection{Identification of $\pi^{\pm}$, $\mathrm{K}^{\pm}$ and $\mathrm{p}$+$\overline{\mathrm{p}}$}
\label{SubSec:Identification}

The particle identification (PID) for pions (\pion), kaons (\kaon) and protons (\proton) used in this analysis is based on a Bayesian technique described in detail in~\cite{Adam:2016acv}, with the time-of-flight $t_{\mathrm{TOF}}$ and the specific energy loss in the \TPC~$\langle \mathrm{d}E/\mathrm{d}x \rangle$ as the input quantities. Different particle species are identified by requiring a minimum probability of 90\%. The PID efficiency of this method is higher than 95\% both for pions and protons up to \pT~$\approx2.5$ \GeV~while for kaons it exhibits a stronger \pT~dependence, reaching 60\% at 2.5 \GeV~with a minimum of 25\% at 4 \GeV. Furthermore, the contamination is below 5\% both for pions and protons, while for kaons it remains below 10\% throughout the entire transverse momentum range considered in this analysis. 

In addition, a different PID procedure that relied on the two-dimensional correlation between the number of standard deviations in units of the resolution from the expected signals of the \TPC~and the \TOF~detectors was also investigated, similar to what was reported in~\cite{Abelev:2014pua}. In this approach particles were selected by requiring their signal to lie within maximum three standard deviations from the $\langle \mathrm{d}E/\mathrm{d}x \rangle$ and $t_{\rm{TOF}}$ values expected for a given particle species and transverse momentum. In addition, the purity was required to be at least 80\%, a condition that becomes essential with increasing transverse momentum where the relevant detector response for different particle species starts to overlap. 

\section{Analysis technique}
\label{Sec:FlowMethods}
In this article, higher flow harmonics for charged pions, charged kaons, protons and anti-protons are reported. In the following paragraphs, the technique used for the measurement of flow harmonics is discussed and an approach to estimate the contribution of non-flow correlations, applied to obtain the final results, is presented. For the estimation of these higher flow harmonics, the symmetry planes are not reconstructed on an event-by-event basis and thus the azimuthal angles of particles are not directly correlated to them. Instead, they are estimated with correlation techniques, where only the azimuthal angles of produced particles are required.

\subsection{Scalar Product method}
In this article, the flow harmonics are calculated with the Scalar Product (SP) method~\cite{Adler:2002pu,Voloshin:2008dg} in which the identified particle of interest (POI) and the charged reference particles (RP) are both selected within the acceptance of the \TPC~detector. This method is based on the calculation of the $Q$-vector from a sample of RP~\cite{Danielewicz:1985hn}, according to 

\begin{equation}
\vec{Q}_{n}=\sum_{k\in{RP}}^{M}e^{in\varphi_{k}},
\label{Qvector}
 \end{equation}
 
where $M$ is the multiplicity of RPs, $\varphi_{k}$ is the azimuthal angle of the $k^{\mathrm{th}}$ reference particle and $n$ is the order of the flow harmonic. 
 
In this study, each event is divided into two subevents ``$a$'' and ``$b$'', covering the ranges $-0.8<\eta<0.0$ and $0.0<\eta<0.8$, respectively. The measured $v_{n}^{a}$ ($v_{n}^{b}$) coefficients are calculated by selecting the identified hadrons (POIs) from subevent ``$a$'' (``$b$'') and the reference particles from subevent ``$b$'' (``$a$'') according to   

\begin{equation}
v_{n}^{a} (p_{\mathrm{T}}) = \frac{\Big \langle{ \Big\langle{\vec{u}_{n}^{k}(p_{\mathrm{T}}) \cdot \frac{\vec{Q}_{n}^{b*}}{M^b}}\Big\rangle}_{k \in a}\Big\rangle}{\sqrt{\Big\langle{\frac{\vec{Q}_{n}^{a}}{M^a}\cdot \frac{\vec{Q}_{n}^{b*}}{M^b}}\Big\rangle}}\,.
\label{vna}
\end{equation}

In Eq.~\ref{vna}, the brackets denote an average over all particles and all events, $M^a$ and $M^b$ are the measured multiplicities of RPs from each subevent in the \TPC~detector, $\vec{u}_{n}^{k} =  e^{in\varphi_{k}}$, $k \in a$, is the unit vector of the $k^{\mathrm{th}}$ POI in subevent ``$a$'', $\vec{Q}_{n}^{a}$ is the $Q$-vector calculated in subevent ``$a$'' and $\vec{Q}_{n}^{b*}$ is the complex conjugate of the $Q$-vector calculated in subevent ``$b$''. The denominator in Eq.~\ref{vna} is referred to further in the text as reference flow. The final measured $v_{n}^{\mathrm{AA}}$~coefficients are calculated as a weighted average of $v_{n}^{a}$ and $v_{n}^{b}$ with the inverse of the square of the statistical uncertainty being the weight. 

The Scalar Product method, used in this article, as well as in~\cite{Abelev:2014pua}, requires less statistics than multi-particle methods, since it is essentially based on two-particle correlations. In addition, it does not introduce any bias originating from multiplicity fluctuations since all $Q$-vectors in Eq.~\ref{vna} are normalised by the relevant multiplicities~\cite{vanderKolk:2012oca}. 

\subsection{Estimation of non-flow correlations}
\label{SubSec:non-flow}

Even after selecting particles from two non-overlapping subevents, a significant residual non-flow contribution remains in the measured flow coefficients. These non-flow contributions are mainly few-particle effects and scale roughly with the inverse of the multiplicity for methods which rely on two-particle correlations, such as the SP. These include correlations originating from jets, resonance decays and quantum statistics correlations which contribute additively to the value of \vnAA. We assume that they do not drastically change with the centrality interval, as discussed in~\cite{Voloshin:2008dg,Voloshin:2007af} and shown in~\cite{Abelev:2014mda}. The corresponding contributions can be estimated using minimum bias pp collisions~\cite{Voloshin:2008dg} and in this article this estimate, denoted as $\delta_n^{AA,pp}$, is subtracted from the measured flow coefficients according to
 
\begin{equation}
v_{n}^{\textrm{sub}}(p_{\mathrm{T}})  = v_{n}^{\textrm{AA}}(p_{\mathrm{T}})   - \delta_{n}^{\textrm{AA,pp}}(p_{\mathrm{T}}) ,
\label{estimated_corrected_vn}
 \end{equation}

\begin{equation}
\delta_{n}^{\textrm{(a)AA,pp}}(p_{\mathrm{T}})  = \frac{\langle{M}\rangle^{\textrm{pp}}\Big\langle{\Big\langle{\vec{u}_{n}^{k}(p_{\mathrm{T}}) \cdot \frac{\vec{Q}^{b*}_{n}}{M^b}}\Big\rangle}_{k \in a}\Big\rangle^{\textrm{pp}}}{\langle{M}\rangle^{\textrm{AA}}\sqrt{\Big\langle{\frac{\vec{Q}_{n}^{a}}{M^a}\cdot \frac{\vec{Q}^{b*}_{n}}{M^b}}\Big\rangle^{\textrm{AA}} } },
\label{deltan}
 \end{equation}
 
\begin{figure}
\begin{center}
\includegraphics[scale=0.82]{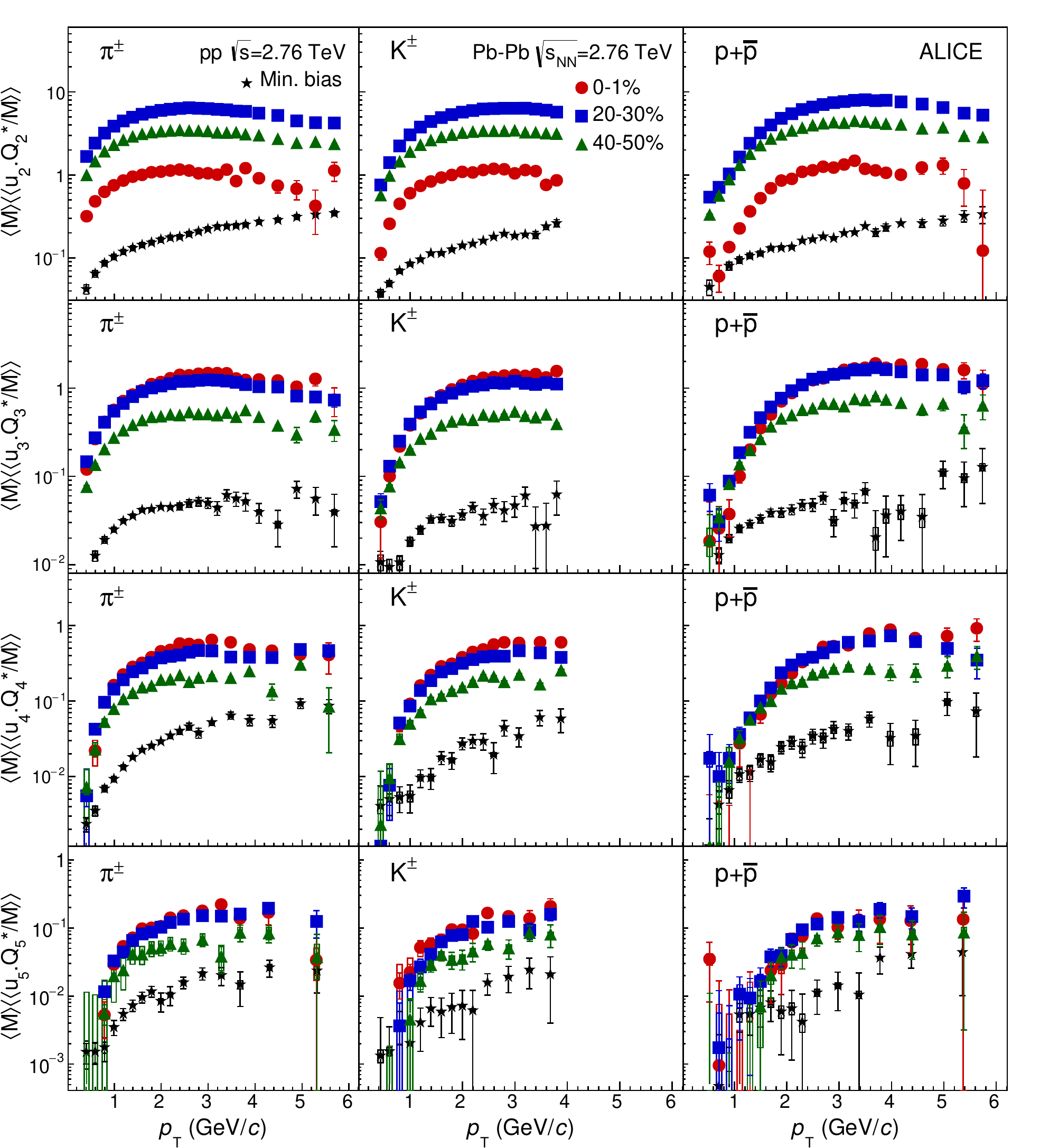}
\end{center}
\caption{The \pT-differential $\langle{M}\rangle\langle{\langle{\vec{u}_{n}\cdot\frac{\vec{Q}^{*}_{n}}{M}}\rangle}\rangle$ of pions (left column), kaons (middle column) and protons (right column) for minimum bias pp and 0--1\%, 20--30\% and 40--50\% centralities in Pb--Pb collisions at \sNN. The rows represent different harmonics.}
\label{uQ_pp}
\end{figure}

where the final $\delta_{n}^{\textrm{AA,pp}}$ is calculated as a weighted average of $\delta_{n}^{\textrm{(a)AA,pp}}$ and $\delta_{n}^{\textrm{(b)AA,pp}}$ with the inverse of the square of the statistical uncertainty as the weight. The term $\delta_{n}^{\textrm{(a)AA,pp}}$ is given by Eq.~\ref{deltan} (similarly for $\delta_{n}^{\textrm{(b)AA,pp}}$). In Eq.~\ref{deltan}, $\langle{M}\rangle^{\textrm{pp}}$ and $\langle{M}\rangle^{\textrm{AA}}$ are the average multiplicities of RPs calculated in pp and Pb--Pb collisions, respectively. In this article, we report the results of $v_{n}^{\textrm{sub}}$, defined in Eq.~\ref{estimated_corrected_vn}, with the superscript `sub' added to stress the applied subtraction procedure. This approach is different compared to previous measurements~\cite{Abelev:2014pua,Abelev:2012di}, where a large pseudorapidity gap $\Delta \eta$ between the POIs and the RPs was used to significantly reduce the contribution from non-flow correlations. The \vtwo~results reported in this article are 2--6\% below the \rawvtwo~measurements reported in \cite{Abelev:2014pua}. This is probably due to the fact that the subtraction procedure using pp collisions accounts for the recoil (away--side) jet which is not accounted for by applying a large $\eta$--gap. On the other hand, it does not account for known medium-induced modifications of jet-like correlations. This could lead to an over-estimation of the non-flow component in high \pT~values. 

Figure~\ref{uQ_pp} presents the \pT-differential $\langle{M}\rangle\langle{\langle{\vec{u}_{n}\cdot\frac{\vec{Q}^{*}_{n}}{M}}\rangle}\rangle$, i.e. the azimuthal correlations scaled by the relevant multiplicities, in pp and Pb--Pb in three centrality intervals (i.e.~0--1\%, 20--30\% and 40--50\%) for all flow harmonics reported in this article for pions, kaons and protons, in the appropriate kinematic range for each species. The data points are drawn with statistical and systematic uncertainties, represented by the error bars and the boxes, respectively. This representation is used in all plots of this article. It is seen that $\langle{M}\rangle^{\textrm{pp}}\langle{\langle{\vec{u}_{2}\cdot\frac{\vec{Q}^{*}_{2}}{M}}\rangle}\rangle^{\textrm{pp}}$ increases monotonically with $p_{\mathrm{T}}$, reaching the magnitude of $\langle{M}\rangle^{\textrm{AA}}\langle{\langle{\vec{u}_{2}\cdot\frac{\vec{Q}^{*}_{2}}{M}}\rangle}\rangle^{\textrm{AA}}$ in ultra-central collisions at high values of \pT, where non-flow correlations are expected to become significant.

\section{Systematic uncertainties}
\label{Sec:Systematics}
The systematic uncertainties are estimated by varying the event and track selection criteria and by studying the detector effects with Monte Carlo (MC) simulations for all particle species, centrality intervals and flow harmonics separately. The contributions from different sources, described below, were extracted from the difference of the \pT-differential \vnAA~(for Pb--Pb collisions) and \uQpp~(for pp collisions) between the default selection criteria described in Section~\ref{Sec:AnalysisDetails} and their variations summarised in Table~\ref{SysVar}. All sources with a statistically significant contribution (i.e.~larger than 3$\sigma$, where $\sigma$ is the uncertainty of the difference between the default results and the ones obtained from the variation of the selection criteria, assuming the two are fully correlated) were then added in quadrature to form the final value of the systematic uncertainty on \vnAA~(or \uQpp) that was propagated to the uncertainty on \vn.

\begin{table}[h!]
\centering
\begin{tabular}{|l|c|c|}
\hline
Error source & Default & Variations\\
\hline
\hline
Primary $z_{vtx}$ & $\pm$10~cm & $\pm$6~cm, $\pm$8~cm\\ 
Centrality estimator  & \VZERO~amplitude & \SPD~clusters, \TPC~tracks\\ 
Magnetic field polarity & both fields & positive, negative \\
Number of \TPC~space points &70& 50, 80, 90, 100\\
$\chi^{2}/\mathrm{ndf}$ per \TPC~space point & 2 & 1, 1.5 \\
$\mathrm{DCA}_{xy}$ ($\mathrm{DCA}_{z}$)~cm &2.4 (3.2)& 0.3, 0.6, 0.9, 1.2\\
Tracking mode & hybrid  & \TPC~standalone, global\\
PID probability & 90\% & 94\%, 98\%\\
MC closure test & --- & --- \\
Non-flow estimate from pp & --- & --- \\
\hline\hline
\end{tabular}
\caption{List of the selection criteria and the corresponding variations used for the estimation of the systematic uncertainties.}
\label{SysVar}
\end{table}

Table~\ref{SystematicsValues} summarises the maximum absolute value, over all transverse momentum and centrality intervals, of the systematic uncertainties from each individual source. These maximum values are obtained for $p_\mathrm{T} > 3$~GeV/$c$ where the typical \vn~values are between 0.1 and 0.2 for \vtwo~(for centrality intervals above the 10--20$\%$ range), 0.07--0.15 for \vthree, 0.05--0.1 for \vfour, and around 0.05 for \vfive, for all sources apart from the DCA variations. In the latter case, the maximum values are obtained for $p_\mathrm{T} < 1$~GeV/$c$, where the \vn~values are significantly smaller.

\begin{table}[h!]
\resizebox{\textwidth}{!}{\begin{tabular}{ |p{4.5cm} ?l|c|c?c|c|c?c|c|c?c|c|c|c| }
\hline
\multicolumn{1}{| c ?}{} & \multicolumn{3}{| c ?}{ \vtwo } & \multicolumn{3}{| c ?}{ \vthree} & \multicolumn{3}{| c ?}{ \vfour} & \multicolumn{3}{| c |}{ \vfive}\\
\hline
Error source  & \pion &  \kaon & \proton &  \pion & \kaon & \proton &  \pion &  \kaon & \proton &  \pion & \kaon &  \proton \\ \hline  \hline
Centrality estimator  & 0.003 & 0.001 & 0.002 & 0.003 & 0.001 & 0.002 & 0.001 & 0.002 & 0.003 & 0.002 & 0.003& 0.006 \\\hline
Magnetic field polarity & \multicolumn{3}{| c ?}{ - } & 0.002 & 0.002 & 0.002 & 0.002 & 0.002 & 0.002 & 0.004 & 0.005 & 0.004 \\\hline
$\mathrm{DCA}_{xy}$ ($\mathrm{DCA}_{z}$) & $10^{-4}$ & -- & $10^{-4}$ & $10^{-4}$ & -- & $10^{-4}$ & $10^{-4}$ & -- & $2\times10^{-4}$ & $10^{-4}$ & -- & $2\times10^{-4}$ \\\hline
Tracking mode  & 0.005 & 0.003 & 0.005 & 0.005 & 0.004 & 0.004 & 0.005 & 0.004 & 0.005 & 0.005 & 0.006 & 0.01 \\\hline
PID probability  & \multicolumn{3}{| c ?}{ - } & \multicolumn{3}{| c ?}{ - } & \multicolumn{3}{| c ?}{ - }& 0.001 & 0.001 & 0.001 \\\hline
MC closure test  & 0.006 & 0.002 & 0.001 & 0.003 & 0.004 & 0.003 & 0.002 & 0.006 & 0.003 & 0.002 & 0.006 & 0.003  \\\hline
Non-flow estimate from pp & \multicolumn{3}{| c ?}{ - }  & \multicolumn{3}{| c ?}{ - }  & 0.001 & 0.001 & 0.003 & 0.001 & 0.001 & 0.003  \\\hline
\hline 
\end{tabular}}
\caption{List of the maximum value of systematic uncertainties from each individual source for each flow harmonic \vn~and particle species. Sources that do not contribute to the systematic uncertainty are not reported in this table.}\label{SystematicsValues}
\end{table}

In order to study the effect of the position of the primary vertex along the beam axis ($z_{vtx}$) on the measurements, the event sample was varied by changing this selection criterium from $\pm$10 cm to $\pm$8 cm and finally to $\pm$6 cm. For all species and centralities, the resulting \vn(\pT) were consistent with results obtained with the default selection. In addition, changing the centrality selection criteria from the signal amplitudes in the \VZERO~scintillator detectors to the multiplicity of \TPC~tracks or the number of \SPD~clusters resulted in maximum contribution of 0.003 (\pion), 0.003 (\kaon), 0.006 (\proton) for all flow harmonics in $p_\mathrm{T} > 3$~GeV/$c$. For $p_\mathrm{T} < 3$~GeV/$c$, the corresponding contributions from this source were significantly smaller in absolute value. Finally, results from runs with different magnetic field polarities did not exhibit any systematic change in \vtwo(\pT) for any particle species or any centrality. For higher harmonics and for $p_\mathrm{T} > 3$~GeV/$c$, the corresponding contributions were at maximum 0.002 for all species and centralities in \vthree(\pT) and \vfour(\pT), and 0.005 in \vfive(\pT), with significantly smaller values for $p_\mathrm{T} < 3$~GeV/$c$.

In addition, the track selection criteria, such as the number of \TPC~space points and the $\chi^2$ per TPC space point per degree of freedom were varied, for all particle species presented in this article. No systematic deviations in the values of \vn(\pT) relative to the results obtained with the default selections were found. The impact of secondary particles on the measured \vn, including products of weak decays, was estimated by varying the selection criteria on both the longitudinal and transverse components of the DCA. This resulted in a non-negligible uncertainty only for pions and anti-protons mainly at low values of transverse momentum (i.e.~$p_\mathrm{T} < 1$~GeV/$c$) as indicated in Table~\ref{SystematicsValues} for all harmonics and centralities. Uncertainties originating from the selected tracking procedure were estimated by using the global or the standalone \TPC~tracking modes (see the discussion in Section~\ref{SubSec:Selection} for details). For all harmonics, differences that contribute to the final systematic uncertainty were found for $p_\mathrm{T} > 3$~GeV/$c$ and their maximum values over all centralities are summarised in Table~\ref{SystematicsValues}. Systematic uncertainties associated with the particle identification procedure were studied by varying the value of the minimum probability of identifying a particle with the Bayesian approach from 90\% to 94\%, and eventually 98\%, but also using an independent technique relying on the number of standard deviations of both the $dE/dx$ ($\sigma_{\mathrm{TPC}}$) and the $t_{\rm{TOF}}$~($\sigma_{\mathrm{TOF}}$) as described in Section~\ref{SubSec:Identification} and in detail in Ref.~\cite{Abelev:2014pua}. These variations did not reveal any systematic differences in the results for \vtwo(\pT), \vthree(\pT) and \vfour(\pT) relative to the results with the default identification requirements. For \vfive(\pT) and for $p_\mathrm{T} > 3$~GeV/$c$ the systematic uncertainty was below 0.001 for all particle species. Systematic uncertainties due to detector inefficiencies were studied using Monte Carlo samples. In particular, the results of the analysis of a sample at the event generator level (i.e.~without invoking either the detector geometry or the reconstruction algorithm) were compared with the results of the analysis over the output of the full reconstruction chain, in a procedure referred to as ``MC closure test''. Table~\ref{SystematicsValues} summarises the maximum contributions over all transverse momenta and centralities, found for $p_\mathrm{T} > 3$~GeV/$c$, for each particle species and harmonic. On the other hand, for $p_\mathrm{T} < 3$~GeV/$c$ the corresponding contributions were significantly smaller.

Furthermore, the contribution from the estimation of non-flow effects extracted with the procedure described in Section~\ref{SubSec:non-flow} was studied by investigating the same list of variations of the event and track selection criteria summarised in Table~\ref{SysVar}, coherently in pp and Pb--Pb collisions. These uncertainties do not account for contributions related to jet quenching effects in Pb--Pb collisions. The maximum differences were negligible for \vtwo(\pT) and \vthree(\pT), and were up to 0.001 for pions and kaons and 0.003 for protons with $p_\mathrm{T} > 3$~GeV/$c$ for \vfour(\pT) and \vfive(\pT).
 
Moreover, the analysis was repeated using different charge combinations (i.e.~positive--positive and negative--negative) for the identified hadrons and the reference particles in both Pb--Pb and pp collisions. The results, after the correction of Eq.~\ref{estimated_corrected_vn}, were compatible with the default ones. Finally, the two subevents used to select POIs and RPs were further separated, by applying a pseudorapidity gap ($\Delta\eta$) between them, from no-gap (default analysis) to $|\Delta\eta| > 0.4$ and eventually reaching $|\Delta\eta| > 0.8$. Both $v_{\textrm{n}}^{\textrm{AA}}$ and \deltan~were calculated using the same gap and the results after the subtraction did not exhibit any systematic change in \vn(\pT) for any particle species or any centrality.

\section{Results and discussion}
\label{Sec:Results}
 
In this section, the results for the \pT-differential \vtwo, \vthree, \vfour~and \vfive~measured in Pb--Pb collisions at \sNN~for 0--1\% up to 40--50\% centrality intervals for pions, kaons and protons are presented. We first present, in Sec.~\ref{Section:CentralityDependece}, the centrality dependence of \vn(\pT) and the relevant contribution of the subtraction terms used to measure \vn. Section~\ref{Section:UltraCentral} focuses on the development of \vn(\pT) for different harmonics in ultra-central collisions. Section~\ref{Section:MassOrdering} presents the mass dependence of \vn(\pT) which is followed by a discussion about the scaling properties of different flow harmonics in different centrality intervals. In Section~\ref{Section:Models}, two models, namely iEBE-VISHNU~\cite{Xu:2016hmp} and A Multi Phase Transport model (AMPT)~\cite{Zhang:1999bd,Lin:2000cx,Lin:2004en}, are compared with the experimental measurements. Note that the same data will be shown in different representations in the following sections to highlight the various physics implications of the measurements. 
 
\subsection{Centrality dependence of flow harmonics}
\label{Section:CentralityDependece}

\begin{figure}[t!]
\begin{center}
\includegraphics[scale=0.72]{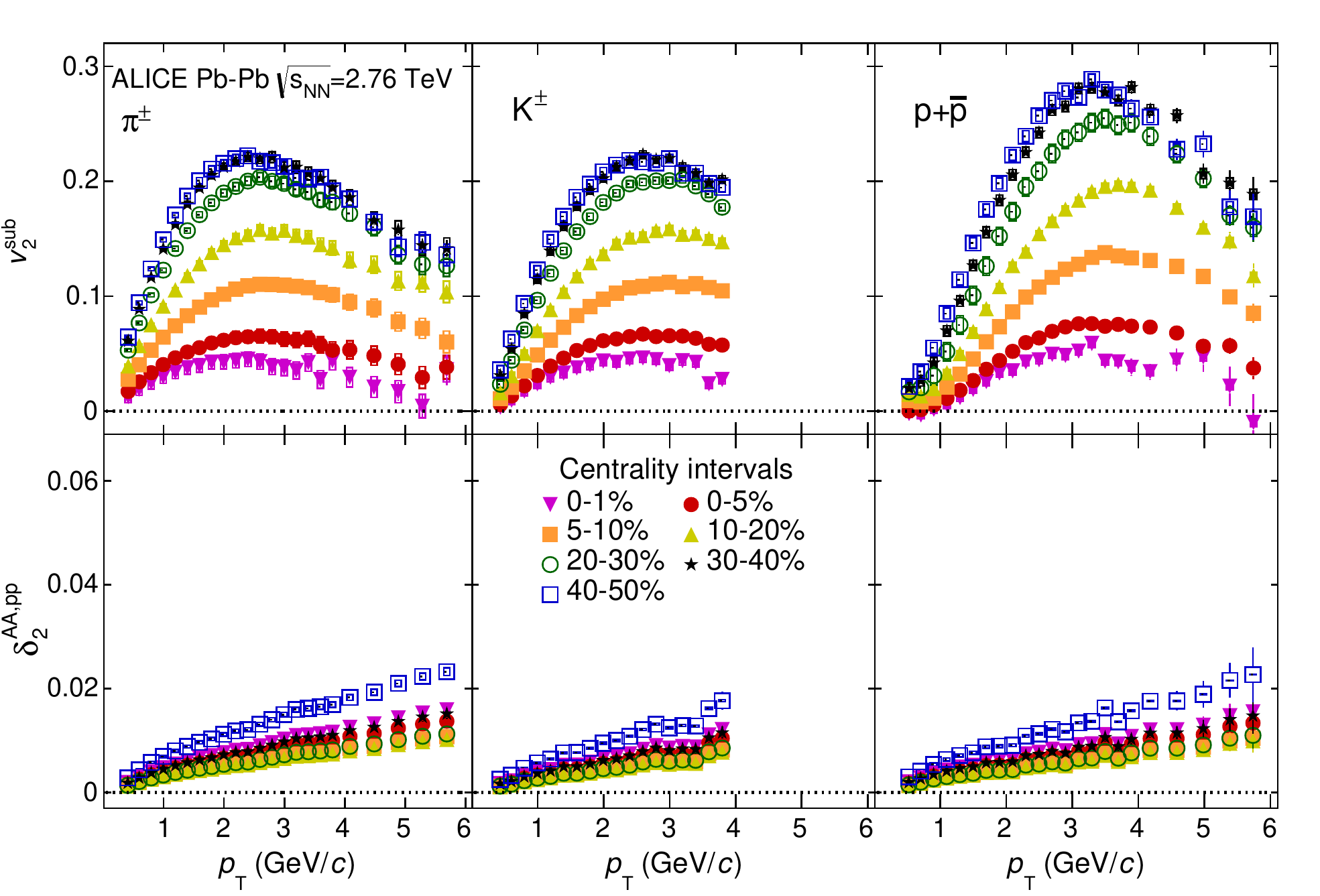}
\end{center}
\caption{The \pT-differential \vtwo~(top row) and \deltatwo~(bottom row) for different centralities in Pb--Pb collisions at \sNN~grouped by particle species.}
\label{v2_centralitydependence}
\end{figure}

Figure~\ref{v2_centralitydependence} presents the \pT-differential \vtwo~(in the top row) and the corresponding subtracted terms denoted as \deltatwo(\pT) (bottom row) for \pion, \kaon~and \proton~measured in different centrality intervals (0--1\% up to 40--50\%) in Pb--Pb collisions at \sNN. The results are grouped in each panel according to particle species to show the dependence of \vtwo(\pT) on centrality. 

This figure illustrates how the value of \vtwo(\pT) increases with centrality (top row) from ultra-central  (0--1\%) to the most peripheral collisions (40--50\%). This is in agreement with the interpretation that the final-state ellipticity of the system originates from the initial-state ellipsoidal geometry in non-central collisions. As illustrated in this figure, this increase for \vtwo(\pT) is smaller for more peripheral collisions: the value of \vtwo(\pT) does not increase significantly from the 30--40\% to the 40--50\% centrality interval despite an increase in the geometrical eccentricity. This feature, which is also observed and discussed in \cite{Abelev:2014pua}, might originate from several effects, such as i) the smaller lifetime of the fireball (the hot, dense and rapidly expanding medium) in peripheral compared to more central collisions that does not allow \vtwo~to develop further, ii) a reduced contribution of eccentricity fluctuations in these centrality intervals compared to more central events or iii) final-state hadronic effects~\cite{Song:2013qma}. In addition, a significant \vtwo(\pT) develops in ultra-central collisions where the collision geometry is almost isotropic and therefore \vtwo~reflects only the contribution from initial-state fluctuations. In summary, the results in Fig.~\ref{v2_centralitydependence} confirm that the geometry of the collision plays a crucial role in the development of \vtwo~as a function of centrality for all particle species. It is also confirmed that the initial-state fluctuations contribute significantly as well.

\begin{figure}[t!]
\begin{center}
\includegraphics[scale=0.72]{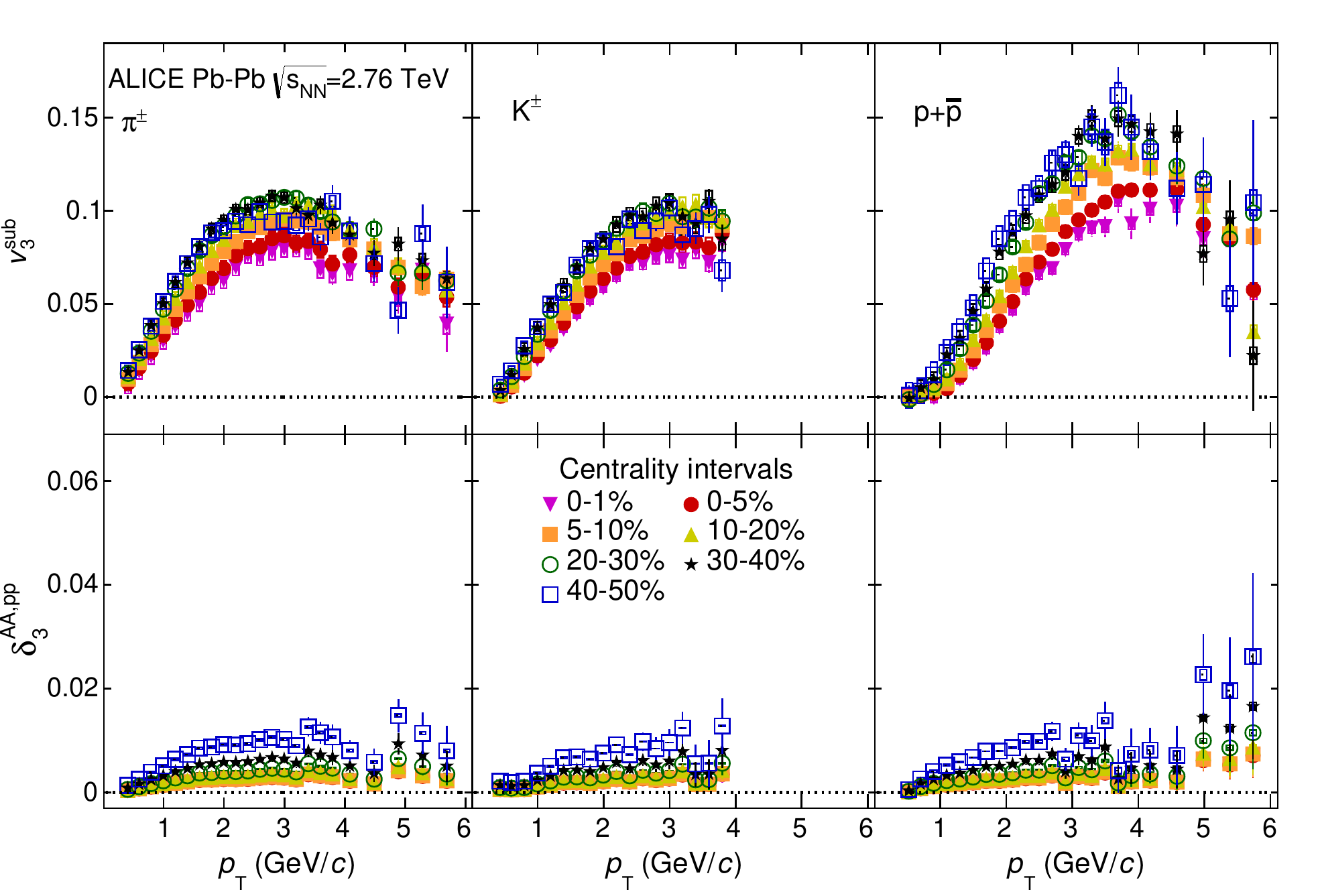}
\end{center}
\caption{The \pT-differential \vthree~(top row) and \deltathree~(bottom row) for different centralities in Pb--Pb collisions at \sNN~grouped by particle species.}
\label{v3_centralitydependence}
\end{figure}
\begin{figure}[th!]
\begin{center}
\includegraphics[scale=0.72]{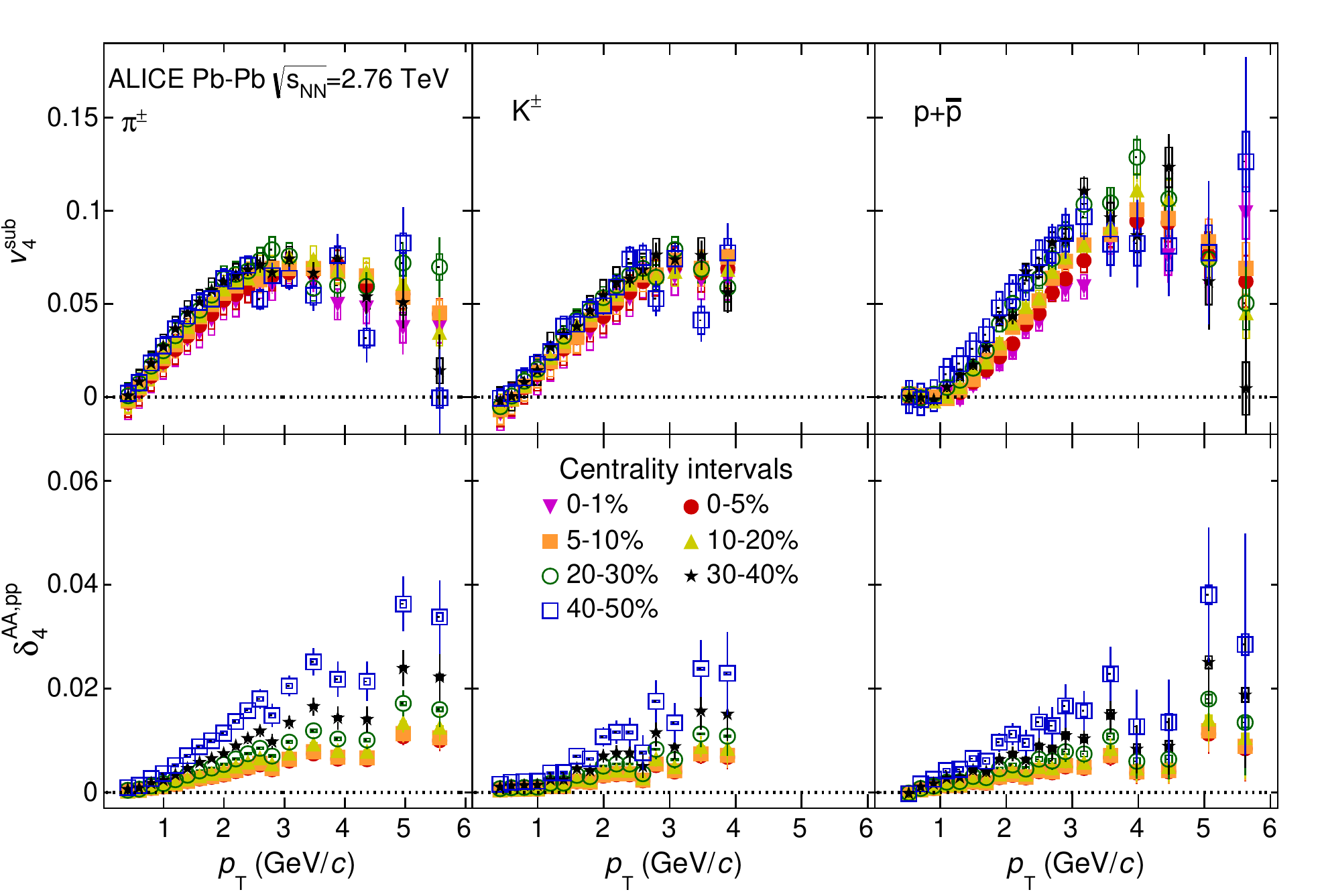}
\end{center}
\caption{The \pT-differential \vfour~(top row) and \deltafour~(bottom row) for different centralities in Pb--Pb collisions at \sNN~grouped by particle species.}
\label{v4_centralitydependence}
\end{figure}

\begin{figure}[th!]
\begin{center}
\includegraphics[scale=0.72]{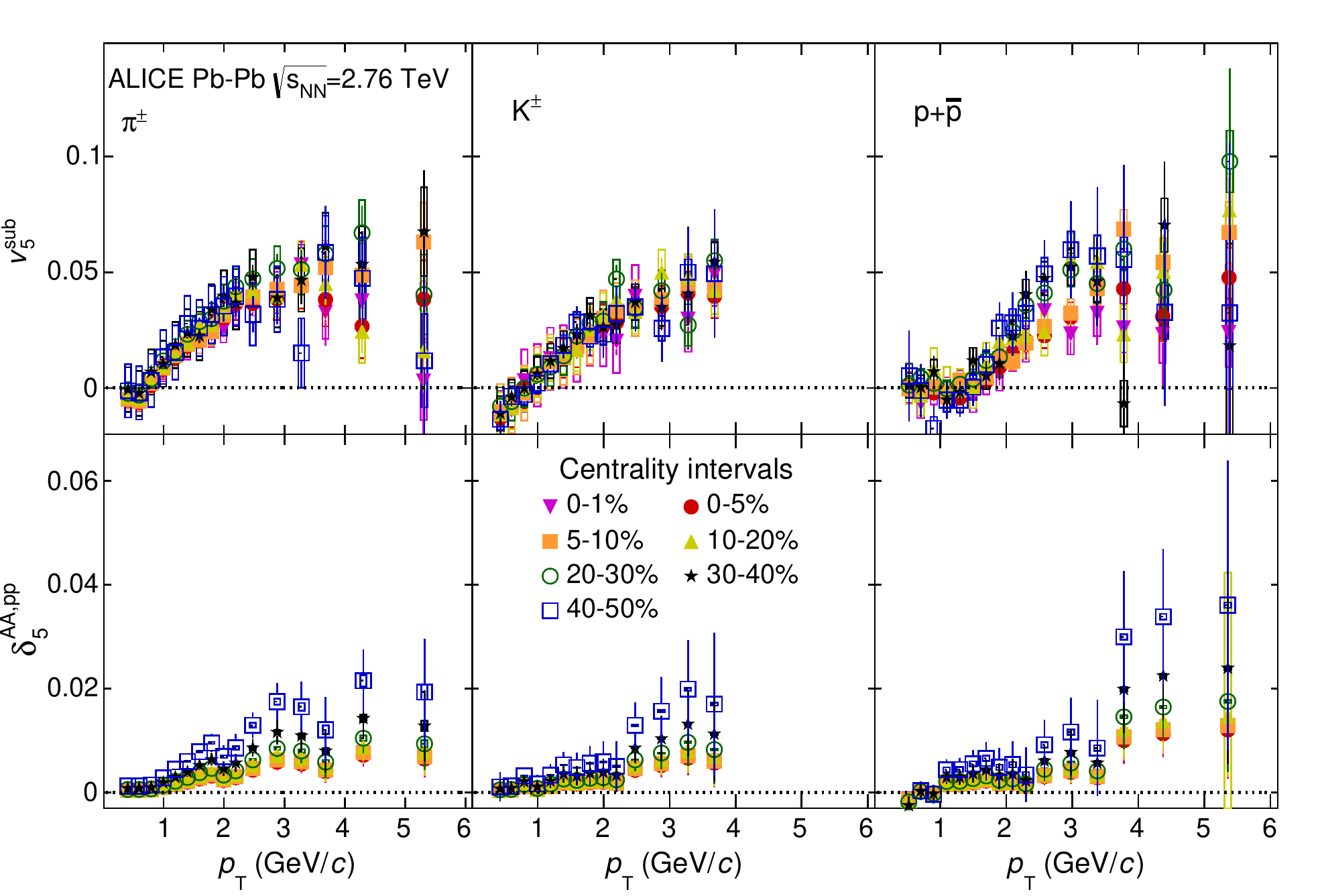}
\end{center}
\caption{The \pT-differential \vfive~(top row) and \deltafive~(bottom row) for different centralities in Pb--Pb collisions at \sNN~grouped by particle species.}
\label{v5_centralitydependence}
\end{figure}

Figure~\ref{v2_centralitydependence} additionally illustrates how \deltatwo~develops with centrality (bottom row). This figure also shows that the value of \deltatwo~becomes larger with increasing transverse momentum, in a \pT~range where non-flow is believed to be a significant contributor to the azimuthal correlations. Furthermore, the relative contribution of \deltatwo~to \vtwoAA~changes as a function of centrality. In particular, the relative value of \deltatwo~is largest for ultra-central collisions (0--1\%) where it is 20\% of \vtwoAA. This percentage drops to 3\% in the 10--20\% centrality interval and increases to 7\% for the most peripheral collisions (40--50\%). This change is also reflected in the absolute value of \deltatwo. The magnitude of \deltatwo~decreases from ultra-central events (0--1\%) to the 10--20\% centrality interval and increases from this centrality interval up to the most peripheral events (40--50\%). This trend as a function of centrality is observed for all particle species and it is due to the interplay between the decrease in multiplicity and the corresponding increase in reference flow as one goes towards more peripheral collisions.

Similar to Fig.~\ref{v2_centralitydependence}, Figs.~\ref{v3_centralitydependence},~\ref{v4_centralitydependence} and~\ref{v5_centralitydependence} present the \pT-differential \vthree, \vfour~and \vfive~(top rows), respectively, and the corresponding subtracted terms (bottom rows) for pions, kaons and protons measured in different centrality intervals. One observes that all \vn~have significant non-zero values throughout the entire measured \pT~range for ultra-central collisions, where the main contributors to the initial coordinate-space anisotropies, which are necessary for the development of \vn, are supposed to be the fluctuations of the initial density profile \cite{ALICE:2011ab}. In addition, the values of the higher flow harmonics increase from ultra-central collisions (0--1\%) to the most peripheral collisions (40--50\%). However, this increase as a function of centrality is smaller in comparison to \vtwo. Thus, \vtwo~seems to mainly reflect the initial geometry of the system while the higher-order flow harmonics are affected less. The non-vanishing values of these higher-order flow harmonics are consistent with the notion in which they are generated primarily from the event-by-event fluctuations of the initial energy density profile.

In addition, all flow harmonics show a monotonic increase with increasing \pT~up to 3 \GeV~reaching a maximum that depends on the particle species and on the collision centrality. In particular, the position of this maximum of \vn(\pT) exhibits a centrality dependence due to the change in radial flow which becomes larger for central compared to peripheral collisions. Moreover, this maximum seems to have a particle mass dependence as well, since it takes place at a higher \pT~value for heavier particles in each centrality interval.

The lower panel of Figs.~\ref{v3_centralitydependence},~\ref{v4_centralitydependence} and~\ref{v5_centralitydependence} also illustrate the magnitude of \deltan~as a function of \pT. In these cases, \deltathree~varies between 5\% and 8\% relative to \vthreeAA, \deltafour~between 12\% and 18\% with respect to \vfourAA, and \deltafive~between 12\% and 20\% with respect to \vfiveAA. Similar to \deltatwo, the variation in the value of higher harmonic \deltan~is derived from the decrease in multiplicity and the increasing reference flow in the transition from central to more peripheral collisions. 

\subsection{Evolution of flow harmonics in ultra-central Pb--Pb collisions}
\label{Section:UltraCentral}

\begin{figure}[!htb]
\begin{center}
\includegraphics[scale=0.82]{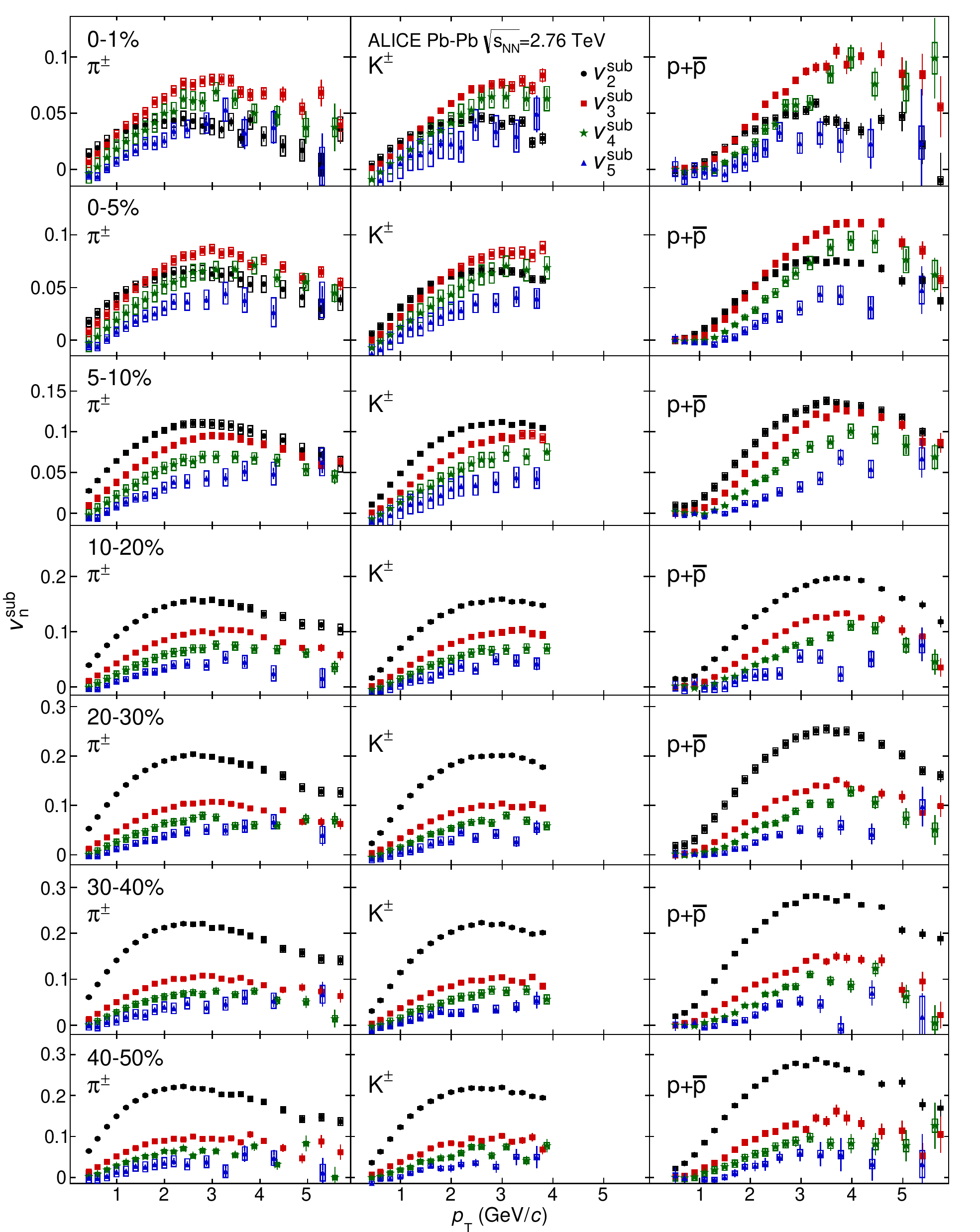}
\end{center}
\caption{The evolution of the \pT-differential \vn~for  \pion, \kaon~and \proton, in the left, middle and right columns, respectively, grouped by centrality interval in Pb--Pb collisions at \sNN.}
\label{Harmonic_Evol}
\end{figure}

Figure~\ref{Harmonic_Evol} shows the evolution of different flow harmonics for \pion~(left column), \kaon~(middle column) and \proton~(right column) for ultra-central (i.e.~0--1\%) collisions in comparison to the other centrality intervals. 

For ultra-central Pb--Pb collisions one expects the influence of the collision geometry to the development of \vn~to be reduced compared to the contribution of initial energy-density fluctuations. Figure~\ref{Harmonic_Evol} shows that for pions the value of \vthree~is equal to \vtwo~at around \pT~$\approx1$ \GeV~and becomes the dominant harmonic for higher transverse momenta. Furthermore, \vfour~at \pT~$\approx2$ \GeV~and \vfive~at around \pT~$\approx3$ \GeV~become equal to \vtwo. For higher transverse momentum values, \vfour~becomes gradually larger than \vtwo~reaching a similar magnitude as \vthree~at around 3.5 \GeV, while \vfive~remains equal to \vtwo. 

As the collisions become more peripheral, one expects that geometry becomes a significant contributor to the development of azimuthal anisotropy. As a result, \vtwo~is the dominant harmonic for peripheral collisions throughout the entire measured momentum range. Furthermore, \vthree, \vfour~and \vfive~seem to have similar magnitudes and \pT~evolution as observed in ultra-central Pb--Pb events, indicating a smaller influence of the collision geometry in their development than for \vtwo.

For kaons and protons, one observes a similar trend in the \pT~evolution of \vtwo, \vthree, \vfour~and \vfive~as for pions. However, the flow harmonics for ultra-central collisions (top middle and right plots of Fig.~\ref{Harmonic_Evol} respectively) exhibit a crossing that takes place at \pT~values that change as a function of the particle mass. For kaons, the crossing between \vtwo~and \vthree~occurs at higher \pT~($\approx$1.4 \GeV) compared to pions while for protons it occurs at an even higher \pT~value ($\approx$1.8 \GeV). Similarly, the \vtwo~and \vfour~crossing occurs higher in \pT~for kaons ($\approx$2.2 \GeV) and protons ($\approx$2.8 \GeV) as compared to pions. The values of \vfour~for kaons reach a similar magnitude to \vthree~at around 3.5 \GeV~and this takes place at around 4 \GeV~for protons. The dependence of the crossing between different flow harmonics, and thus the range where a given harmonic becomes dominant, on the particle mass can be attributed to the interplay of not only elliptic but also triangular and quadrangular flow with radial flow. 

\begin{figure}[t!]
\begin{center}
\includegraphics[width=0.49\textwidth]{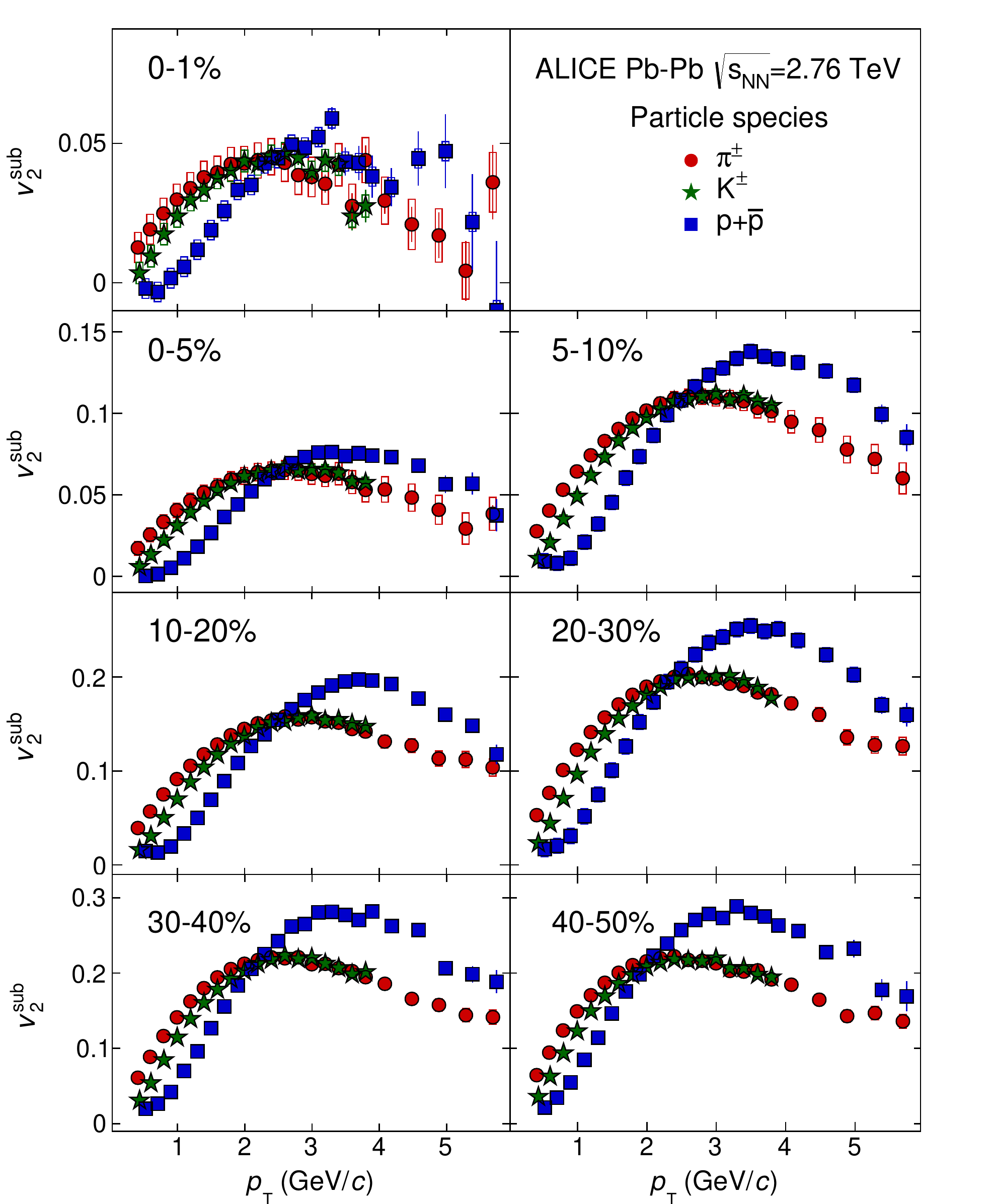}
\includegraphics[width=0.49\textwidth]{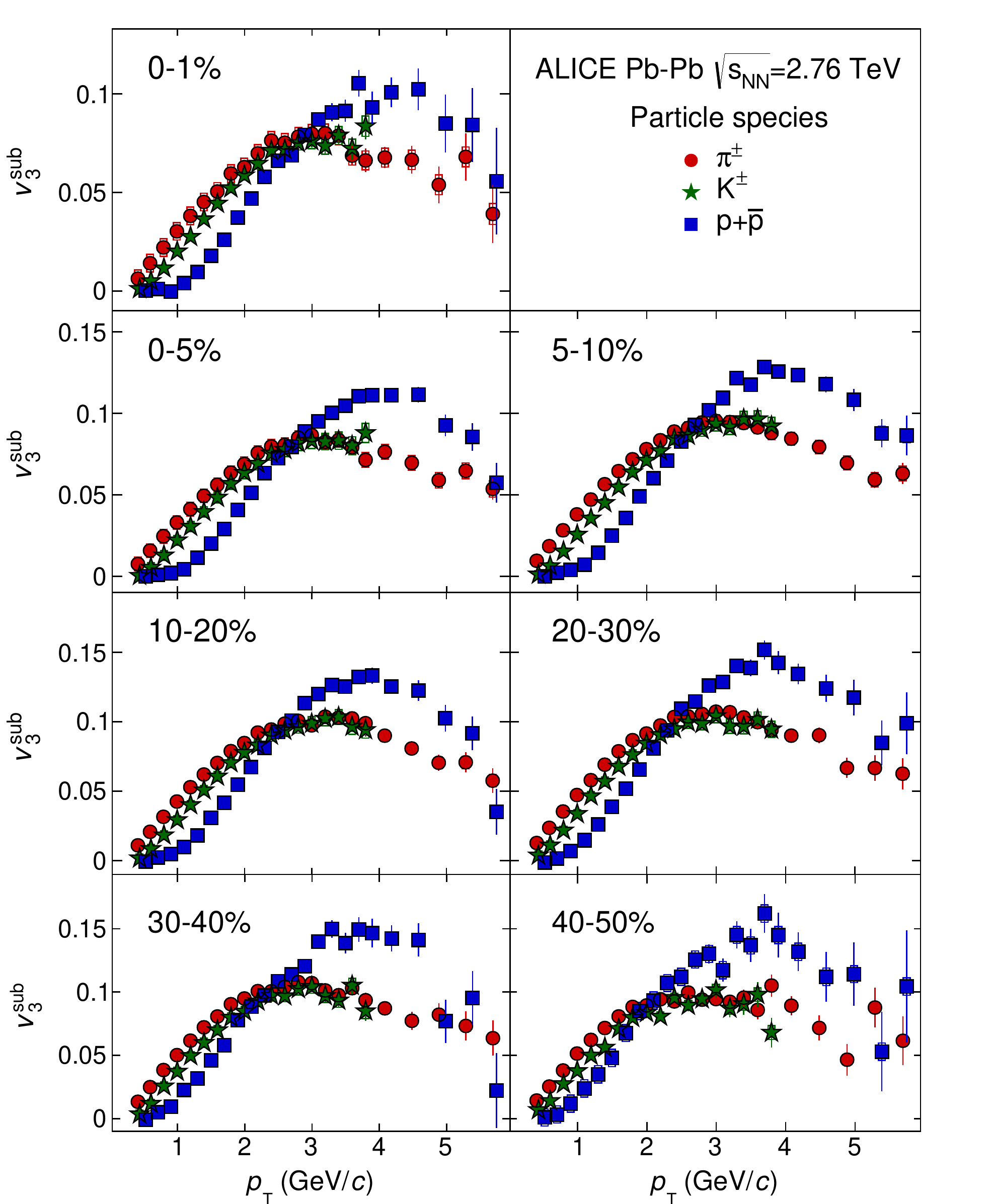}
\end{center}
\captionof{figure}{The \pT-differential \vtwo~(left figure) and \vthree~(right figure)~for different particle species grouped by centrality class in Pb--Pb collisions at \sNN.}
\label{v2v3_mass}
\end{figure}
\begin{figure}[th!]
\begin{center}
\includegraphics[width=0.49\textwidth]{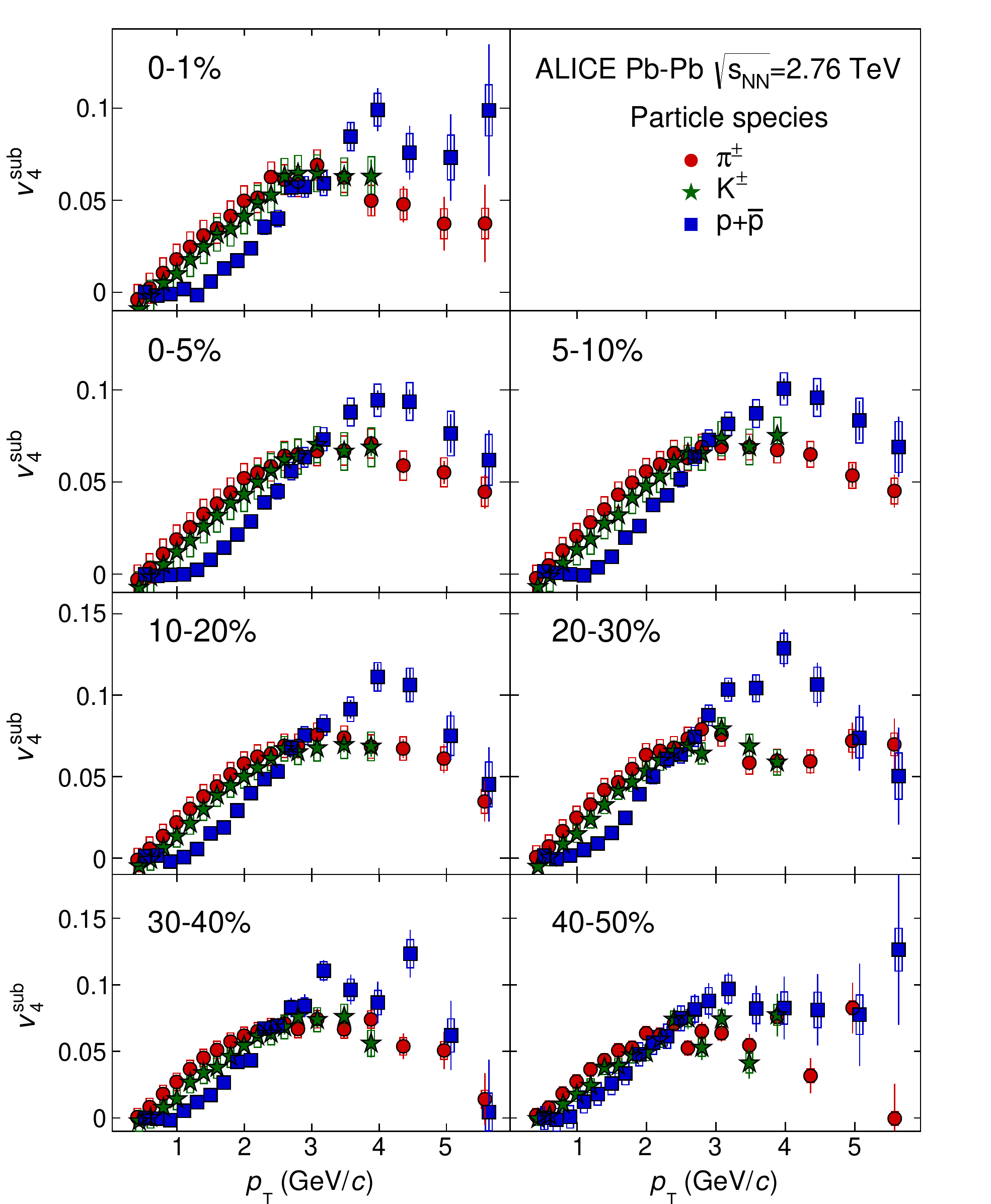}
\includegraphics[width=0.49\textwidth]{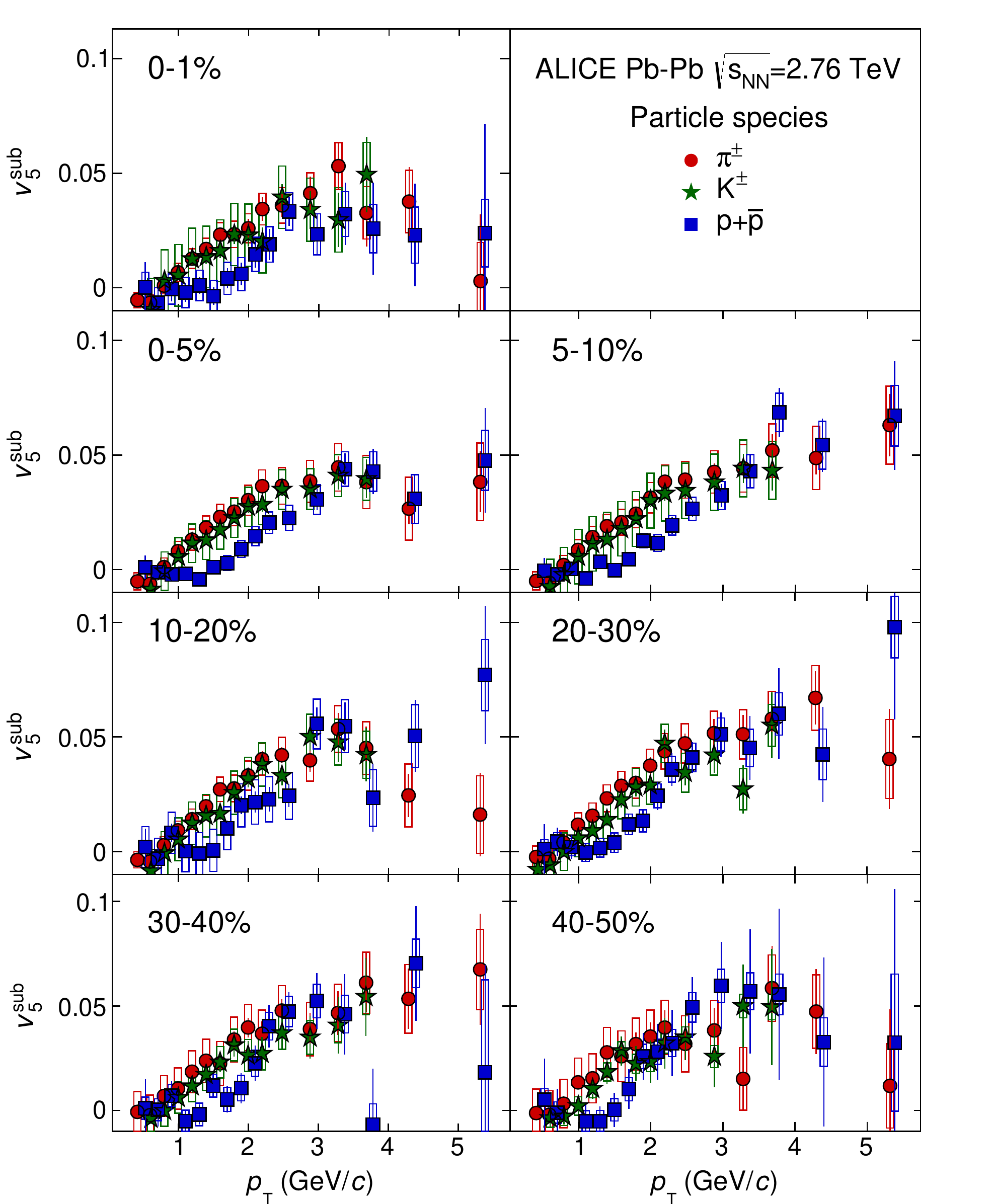}
\end{center}
\captionof{figure}{The \pT-differential \vfour~(left figure) and \vfive~(right figure)~for different particle species grouped by centrality class in Pb--Pb collisions at \sNN.}
\label{v4v5_mass}
\end{figure}

\subsection{Mass ordering}
\label{Section:MassOrdering}

The interplay between the different flow harmonics and radial flow can be further probed by studying how \vn(\pT) develops as a function of the particle mass for various centralities. In Ref.~\cite{Abelev:2014pua}, it was clearly demonstrated that the interplay between radial and elliptic flow leads to a characteristic mass ordering at \pT~$<$ 2--3 \GeV. This mass ordering originates from the fact that radial flow creates a depletion in the particle spectrum at low \pT~values, which increases with increasing particle mass and transverse velocity. When this effect is embedded in an environment where azimuthal anisotropy develops, it leads to heavier particles having smaller \vn~values compared to lighter ones at given values of \pT. It is thus interesting to study whether the interplay between the anisotropic flow harmonics and radial flow leads also to a mass ordering in \vn(\pT) for $n > 2$.

Figure~\ref{v2v3_mass}--left presents the \pT-differential \vtwo~for charged pions, kaons and protons starting from ultra-central collisions up to the 40--50\% centrality interval. The observed evolution of \vtwo~with mass confirms that the interplay between elliptic and radial flow leads to lower \vtwo~values at fixed \pT~for heavier particles for \pT~$<$~2--3 \GeV, depending on the centrality interval. 

Similarly, Figs.~\ref{v2v3_mass}--right,~\ref{v4v5_mass}--left and~\ref{v4v5_mass}--right show the \pT-differential \vthree, \vfour~and \vfive, respectively, for different particle species and for each centrality interval. A clear mass ordering is seen in the low \pT~region, i.e.~for \pT~$<$ 2--3 \GeV, for \vthree(\pT), \vfour(\pT) and \vfive(\pT), which arises from the interplay between the anisotropic flow harmonics and radial flow. 

Furthermore, the \vn(\pT) values show a crossing between pions, kaons and protons, that, depending on the centrality and the order of the flow harmonic, takes place at different \pT~values. In Figs.~\ref{v2v3_mass} and~\ref{v4v5_mass} it is seen that the crossing between, e.g.~\pion~and \proton~occurs at lower \pT~for more peripheral collisions in comparison to more central events. The crossing point for central collisions occurs at higher \pT~values for \vn~since the common velocity field, which exhibits a significant centrality dependence, affects heavy particles more. The current study shows that this occurs not only in the case of elliptic flow but also for higher flow harmonics. Finally, beyond the crossing point for each centrality and for every harmonic, it is seen that particles tend to group based on their number of constituent quarks. This apparent grouping will be discussed in the next subsection.

\begin{figure}[t]
\begin{center}
\includegraphics[width=0.49\textwidth]{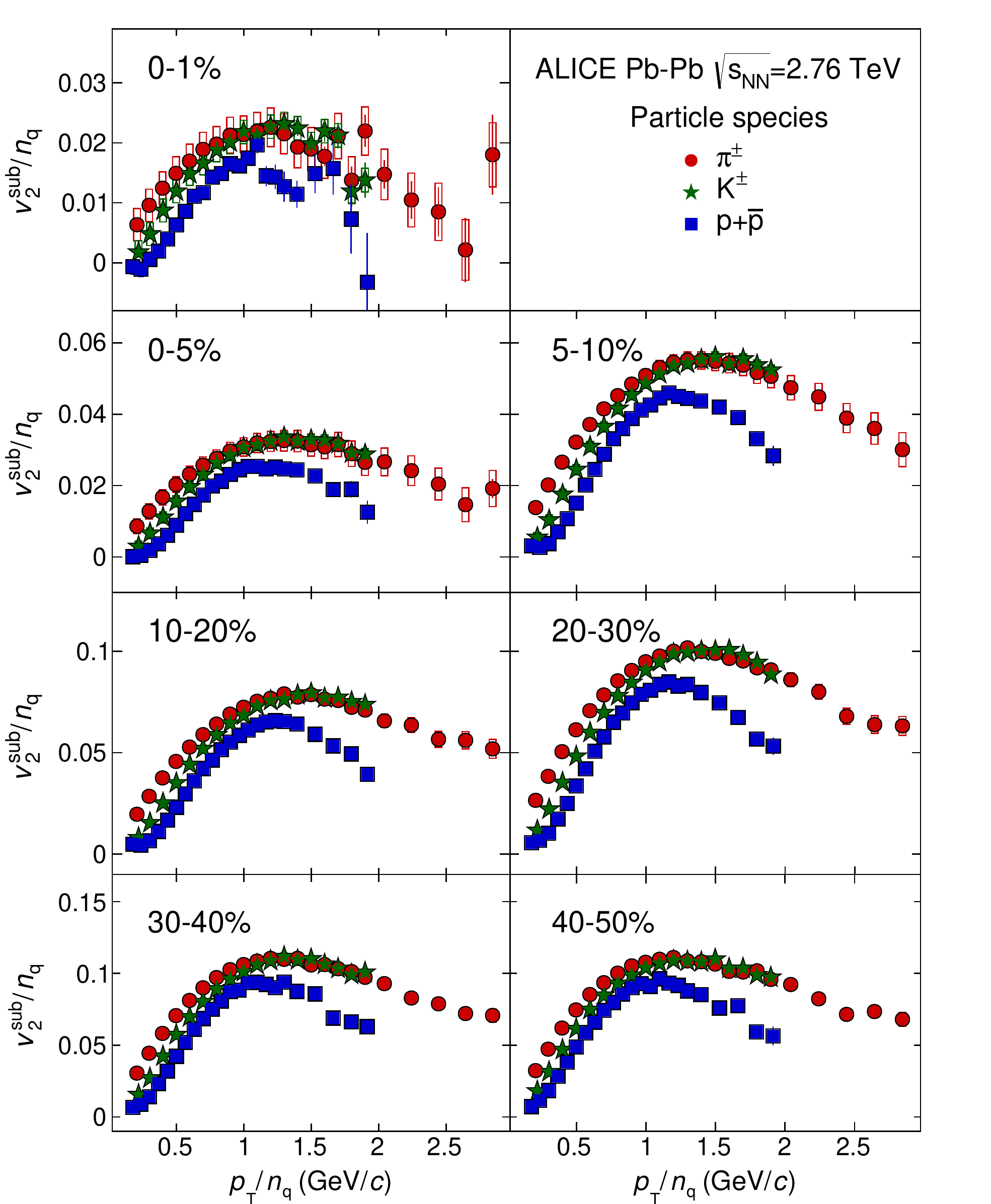}
\includegraphics[width=0.49\textwidth]{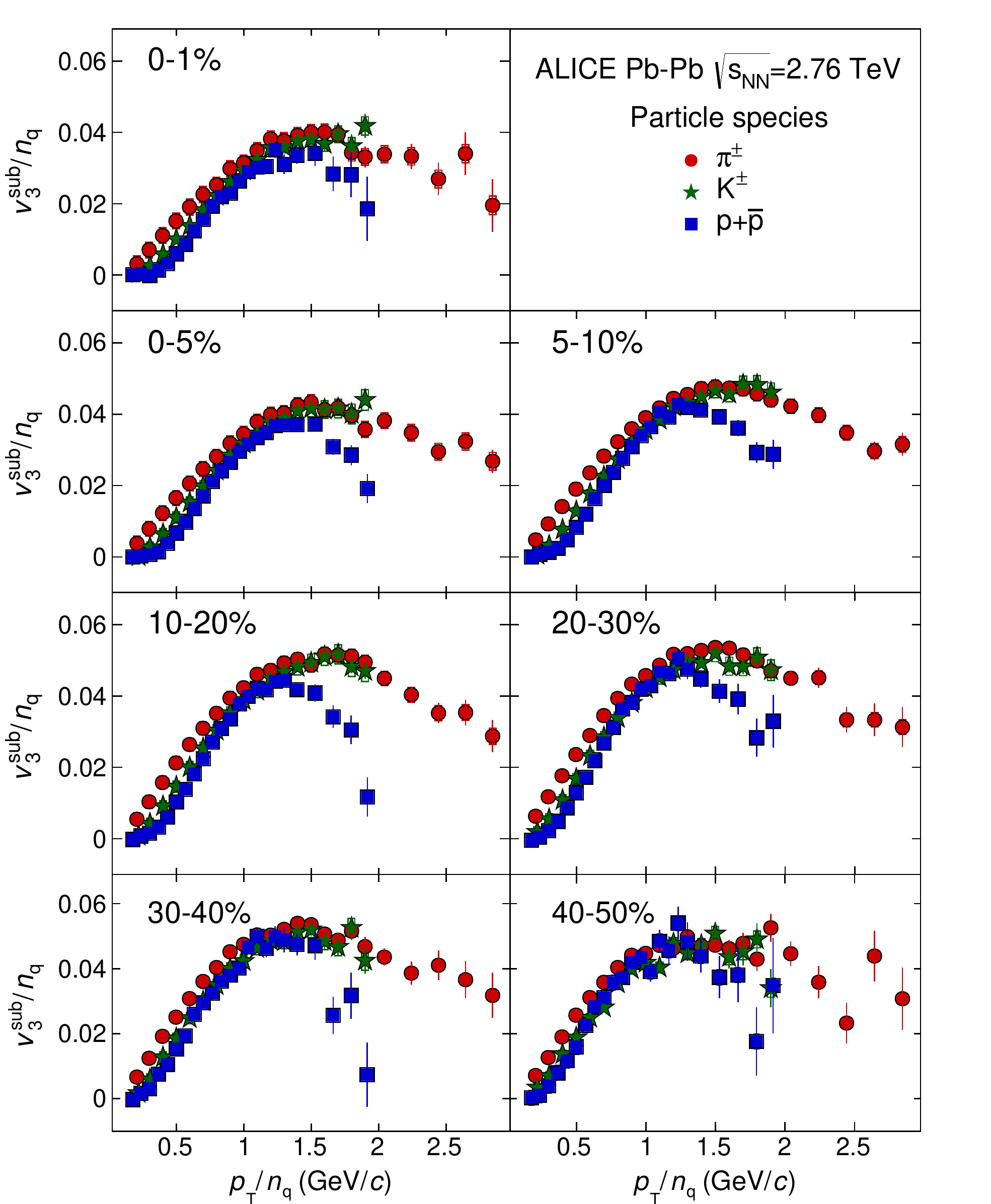}
\end{center}
\captionof{figure}{The \pTnq~dependence of \vtwonq~(left figure)~and \vthreenq~(right figure)~for \pion, \kaon and \proton~for Pb--Pb collisions in various centrality intervals at \sNN.}
\label{v2v3_NCQ}
\end{figure}

\begin{figure}[th!]
\begin{center}
\includegraphics[width=0.49\textwidth]{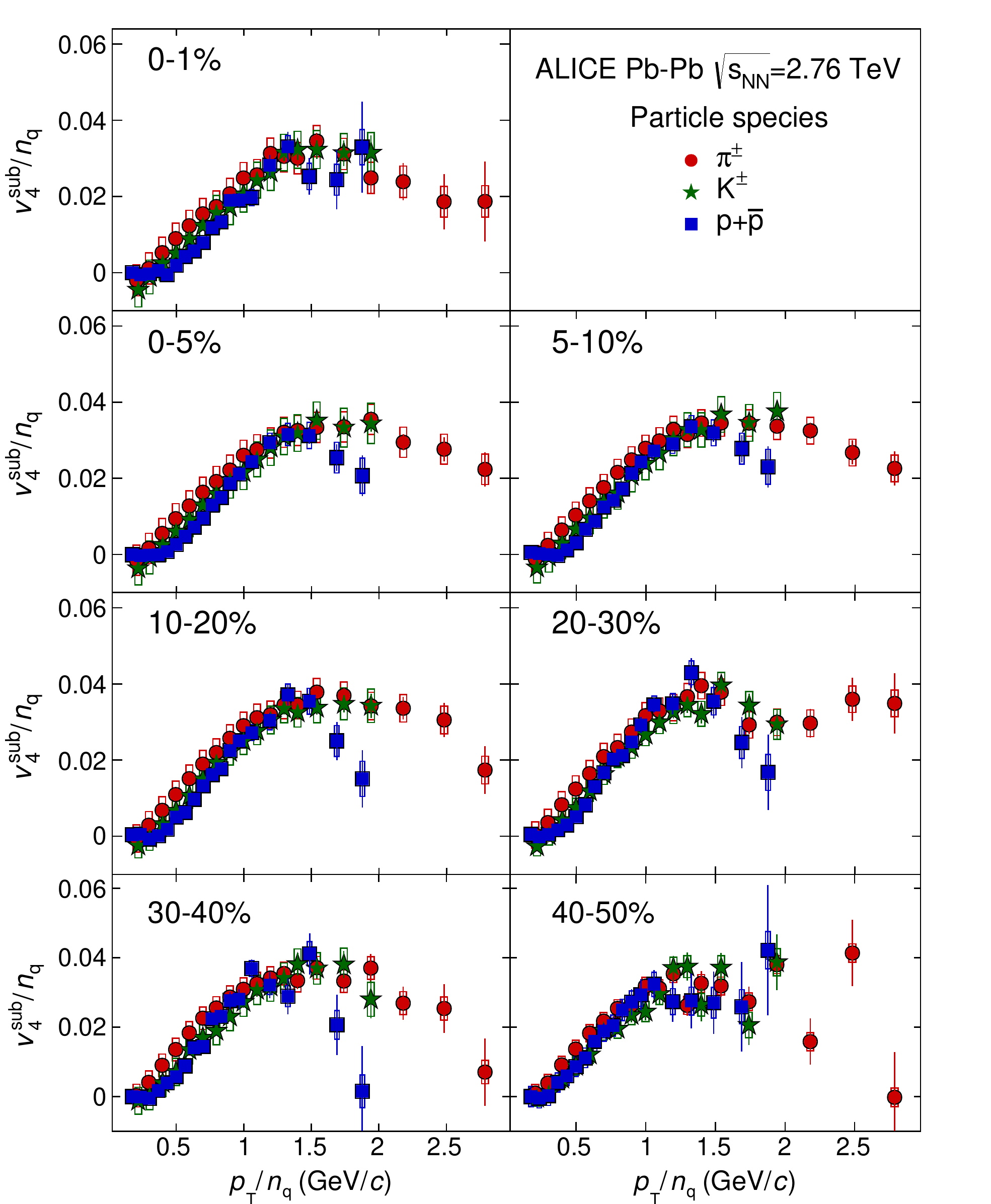}
\includegraphics[width=0.49\textwidth]{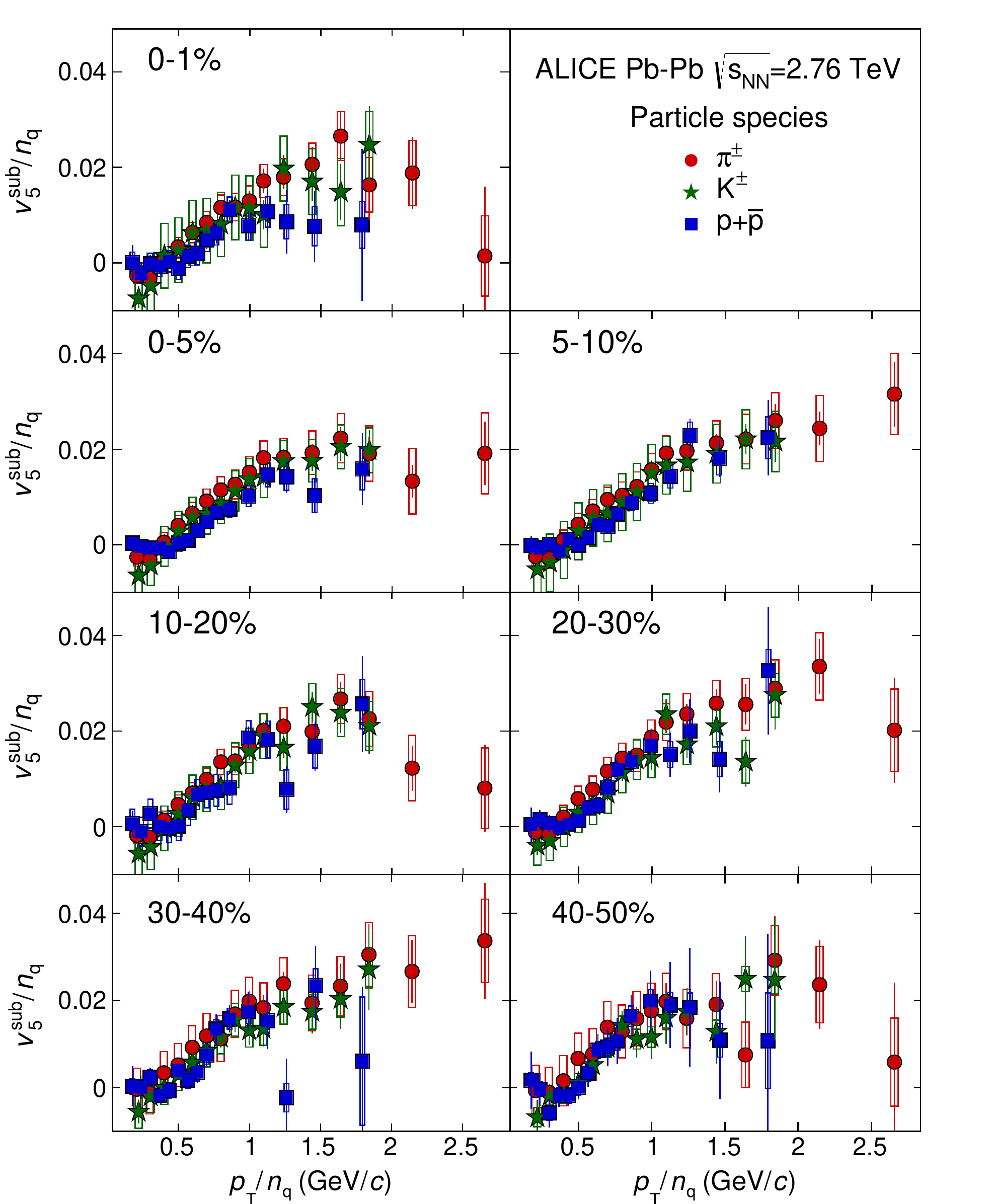}
\end{center}
\captionof{figure}{The \pTnq~dependence of \vfournq~(left figure)~and \vfivenq~(right figure)~for \pion, \kaon and \proton~for Pb--Pb collisions in various centrality intervals at \sNN.}
\label{v4v5_NCQ}
\end{figure}

\subsection{Test of scaling properties}
\label{Section:Scaling}

It was first observed at RHIC that at intermediate values of transverse momentum ($3 <$ \pT~$< 6$  \GeV) the value of $v_2$ for baryons is larger than that of mesons~\cite{Adams:2003am, Abelev:2007qg, Adler:2003kt,  Adare:2006ti}. As a result it was suggested that if both \rawvn~and \pT~are scaled by the number of constituent quarks (\nq), the resulting $p_{\mathrm{T}}/n_{q}$ dependence of the scaled values for all particle species will have an approximate similar magnitude and dependence on scaled transverse momentum. This scaling, known as number of constituent quark scaling (NCQ), worked fairly well at RHIC energies, although later measurements revealed sizeable deviations from a perfect scaling~\cite{Adare:2012vq}. Recently, ALICE measurements~\cite{Abelev:2014pua} showed that the NCQ scaling at LHC energies holds at an approximate level of $\pm$20\% for \vtwoAA.

Although the scaling is only approximate, it stimulated various theoretical ideas that attempted to address its origin. As a result, several models~\cite{Voloshin:2002wa,Molnar:2003ff} attempted to explain this observed effect by requiring quark coalescence to be the dominant particle production mechanism in the intermediate \pT~region, where the hydrodynamic evolution of the fireball is not the driving force behind the development of anisotropic flow. 

Figures~\ref{v2v3_NCQ} and~\ref{v4v5_NCQ} present \vtwo~and~\vthree, as well as \vfour~and \vfive, respectively, scaled by the number of constituent quarks (\nq) as a function of \pTnq~for \pion, \kaon~and \proton~grouped in centrality bins. Figure~\ref{v2v3_NCQ}--left is consistent with the observation reported in~\cite{Abelev:2014pua} related to the elliptic flow. For higher harmonics this scaling holds at the same level ($\pm$20\%) within the current statistical and systematic uncertainties. 

\subsection{Comparison with models}
\label{Section:Models}

Measurements of \vnAA~at RHIC and LHC have been successfully described by hydrodynamical calculations. In particular in~\cite{Abelev:2014pua} it was shown that a hybrid model that couples the hydrodynamical expansion of the fireball to a hadronic cascade model describing the final-state hadronic interactions is able to reproduce the basic features of the measurements at low values of \pT. In parallel, various other models that incorporate a different description of the dynamical evolution of the system, such as AMPT, are also able to describe some of the main features of measurements of azimuthal anisotropy~\cite{Zhang:1999bd,Lin:2000cx,Lin:2004en}. In this section, these two different theoretical approaches will be confronted with the experimental measurements.

\subsubsection{Comparison with iEBE-VISHNU}
\label{Section:Hydro}

\begin{figure}[t!]
\begin{center}
\includegraphics[scale=0.65]{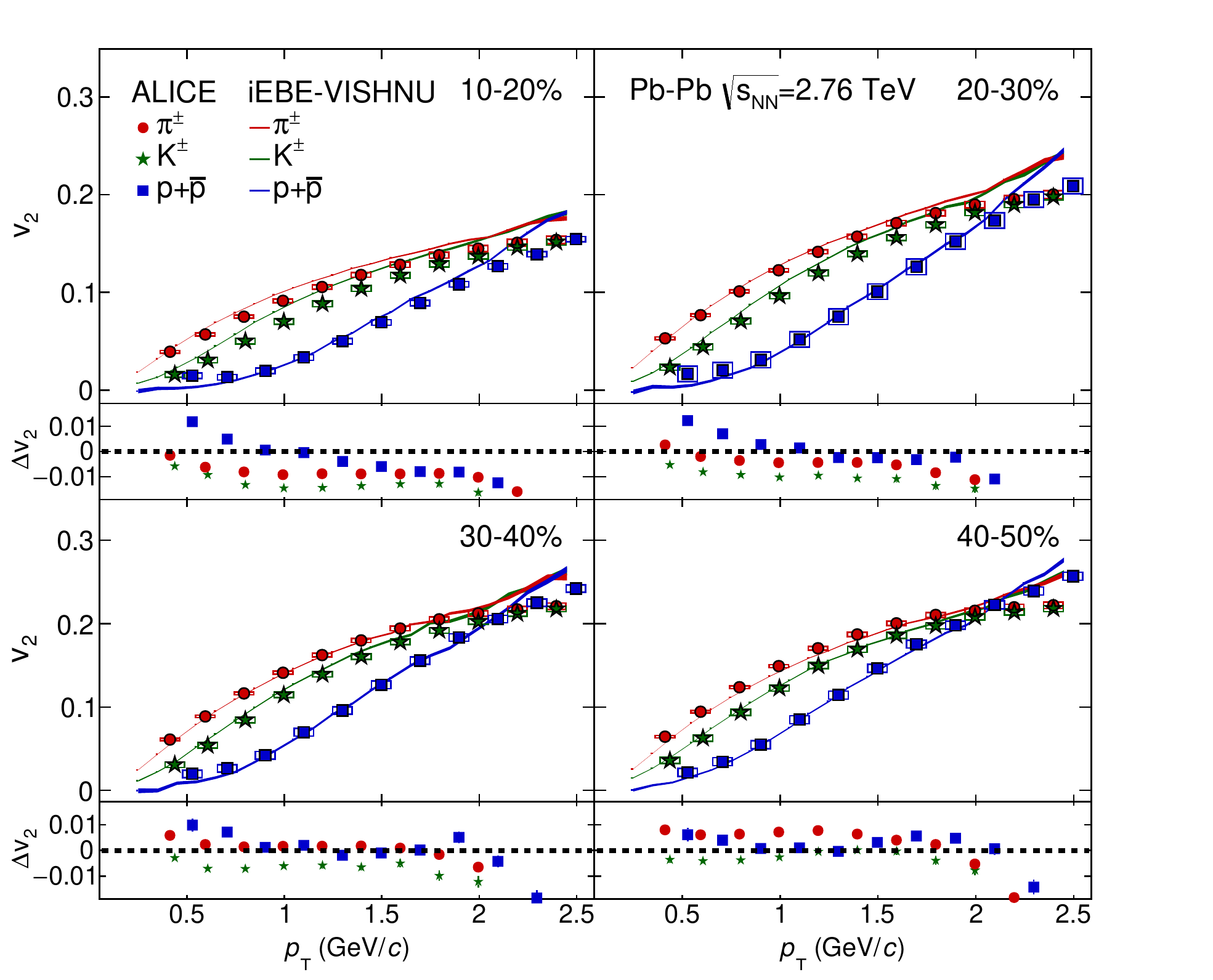}
\end{center}
\captionof{figure}{The \pT-differential \vtwo~for pions, kaons and protons measured with the Scalar Product method in Pb--Pb collisions at \sNN~compared to \rawvtwo~measured with iEBE-VISHNU. The upper panels present the comparison for 10--20\% up to 40--50\% centrality intervals. The thickness of the curves reflect the uncertainties of the hydrodynamical calculations. The differences between \vtwo~from data and \rawvtwo~from iEBE-VISHNU are presented in the lower panels.}
\label{hydro_v2}
\end{figure}

\begin{figure}[t!]
\begin{center}
\includegraphics[scale=0.65]{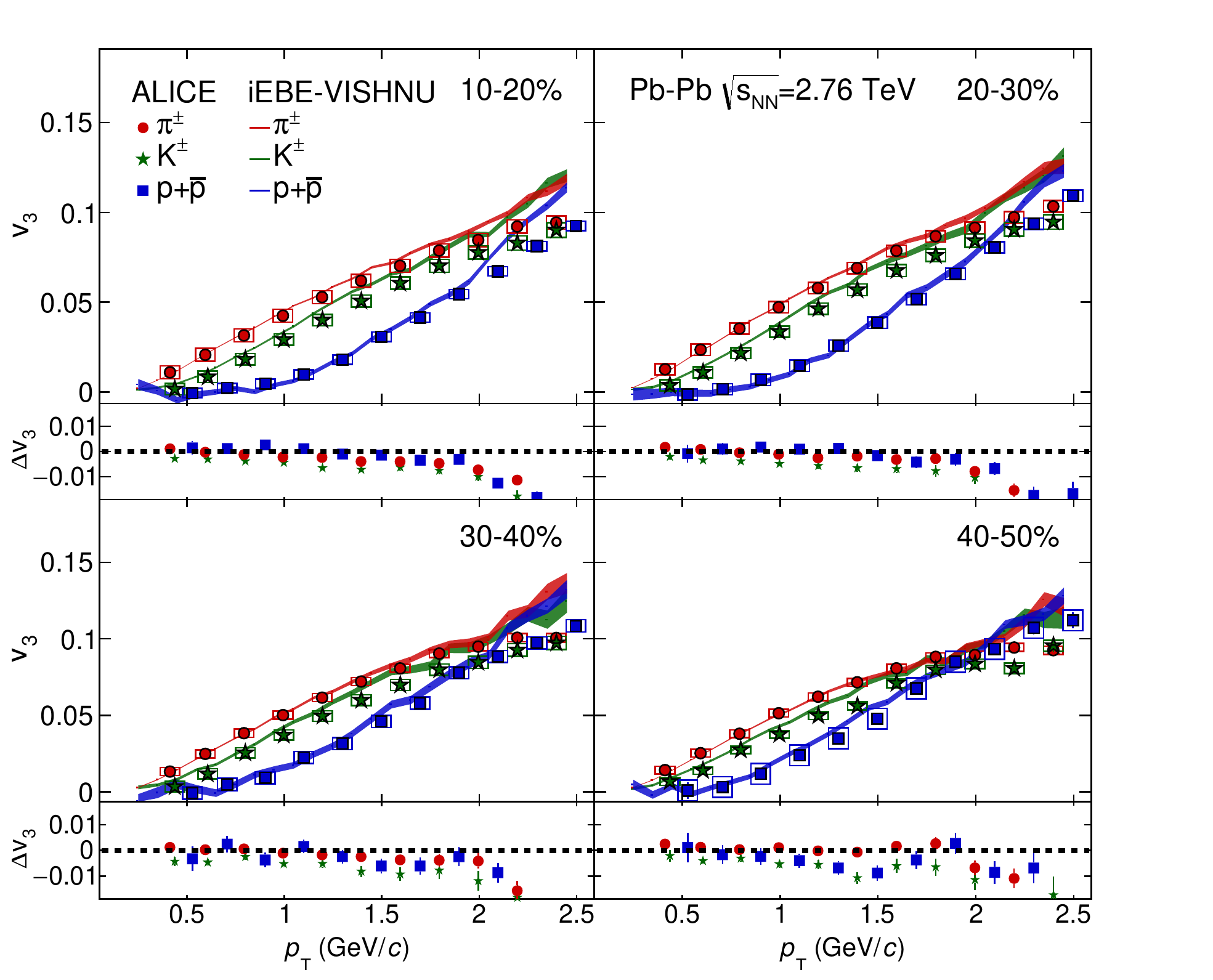}
\end{center}
\captionof{figure}{The \pT-differential \vthree~for pions, kaons and protons measured with the Scalar Product method in Pb--Pb collisions at \sNN~compared to \rawvthree~measured with iEBE-VISHNU. The upper panels present the comparison for 10--20\% up to 40--50\% centrality intervals. The thickness of the curves reflect the uncertainties of the hydrodynamical calculations. The differences between \vthree~from data and \rawvthree~from iEBE-VISHNU are presented in the lower panels.}
\label{hydro_v3}
\end{figure}

\begin{figure}[th!]
\begin{center}
\includegraphics[scale=0.65]{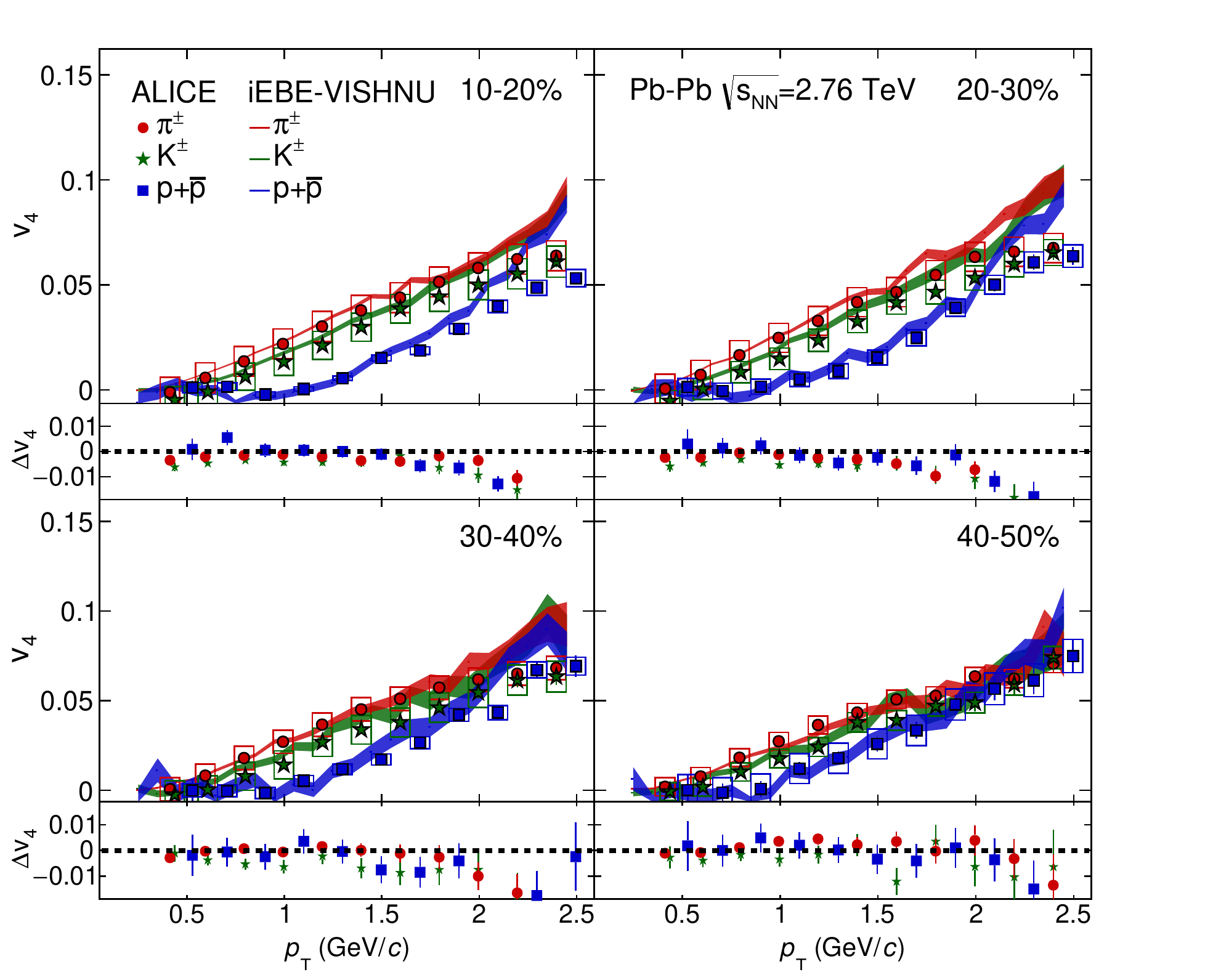}
\end{center}
\captionof{figure}{The \pT-differential \vfour~for pions, kaons and protons measured with the Scalar Product method in Pb--Pb collisions at \sNN~compared to \rawvfour measured with iEBE-VISHNU. The upper panels present the comparison for 10--20\% up to 40--50\% centrality intervals. The thickness of the curves reflect the uncertainties of the hydrodynamical calculations. The differences between \vfour~from data and \rawvfour~from iEBE-VISHNU are presented in the lower panels.}
\label{hydro_v4}
\end{figure}

Figures~\ref{hydro_v2},~\ref{hydro_v3} and~\ref{hydro_v4} present the comparison between the ALICE measurements of \vn~and recent \rawvn~hydrodynamical calculations from~\cite{Xu:2016hmp}. These calculations are based on iEBE-VISHNU, an event-by-event version of the VISHNU hybrid model~\cite{Shen201661} which couples 2+1 dimensional viscous hydrodynamics (VISH2+1) to a hadron cascade model (UrQMD)~\cite{Bass:1998ca} and uses a set of fluctuating initial conditions generated with AMPT. The iEBE-VISHNU model makes it possible to study the influence of the hadronic stage on the development of elliptic flow and higher harmonics for different particles. In this model, the initial time after which the hydrodynamic evolution begins is set to $\tau_{0} = 0.4$~$\mathrm{fm}/c$ and the transition between the macroscopic and microscopic approaches takes place at a temperature of $T = 165$~MeV. Finally, the value of the shear viscosity to entropy density ratio is chosen to be $\eta/s = 0.08$, corresponding to the conjectured lower limit discussed in the introduction. These input parameters were chosen to best fit the multiplicity and transverse momentum spectra of charged particles in most central Pb--Pb collisions as well as the $p_{\mathrm{T}}$-differential $v_2$, $v_3$, and $v_4$ for charged particles for various centrality intervals.
  
These figures show that this hydrodynamical calculation can reproduce the observed mass ordering in the experimental data for pions, kaons and protons. In particular, it is seen that for the range $1<$~\pT~$<2$ \GeV~in the 10--20\% centrality interval the model overpredicts the pion \vtwo(\pT) values by an average of 10\%, however for more peripheral collisions the curve describes the data points relatively well. In addition, the model describes \vthree~and \vfour~for charged pions within 5\%, i.e.~better than \vtwo. Furthermore, it is seen that iEBE-VISHNU overpredicts the \vtwo(\pT) values of \kaon~(i.e.~10--15\% deviations) and does not describe \proton~in more central collisions (i.e.~by 10\% with a different transverse momentum dependence compared to data), but in more peripheral collisions the agreement with the data points is better. Finally, the model describes the \vthree(\pT) and \vfour(\pT) values for \kaon~and \proton~with a reasonable accuracy (i.e.~within 5\%) in all centrality intervals up to \pT~around 2 \GeV. These observations are also illustrated in the lower plots of each panel in Figs.~\ref{hydro_v2},~\ref{hydro_v3} and~\ref{hydro_v4} that present the difference between the measured \vn~relative to a fit to the hydrodynamical calculation. 

\begin{figure}[t!]
\begin{center}
\includegraphics[scale=0.65]{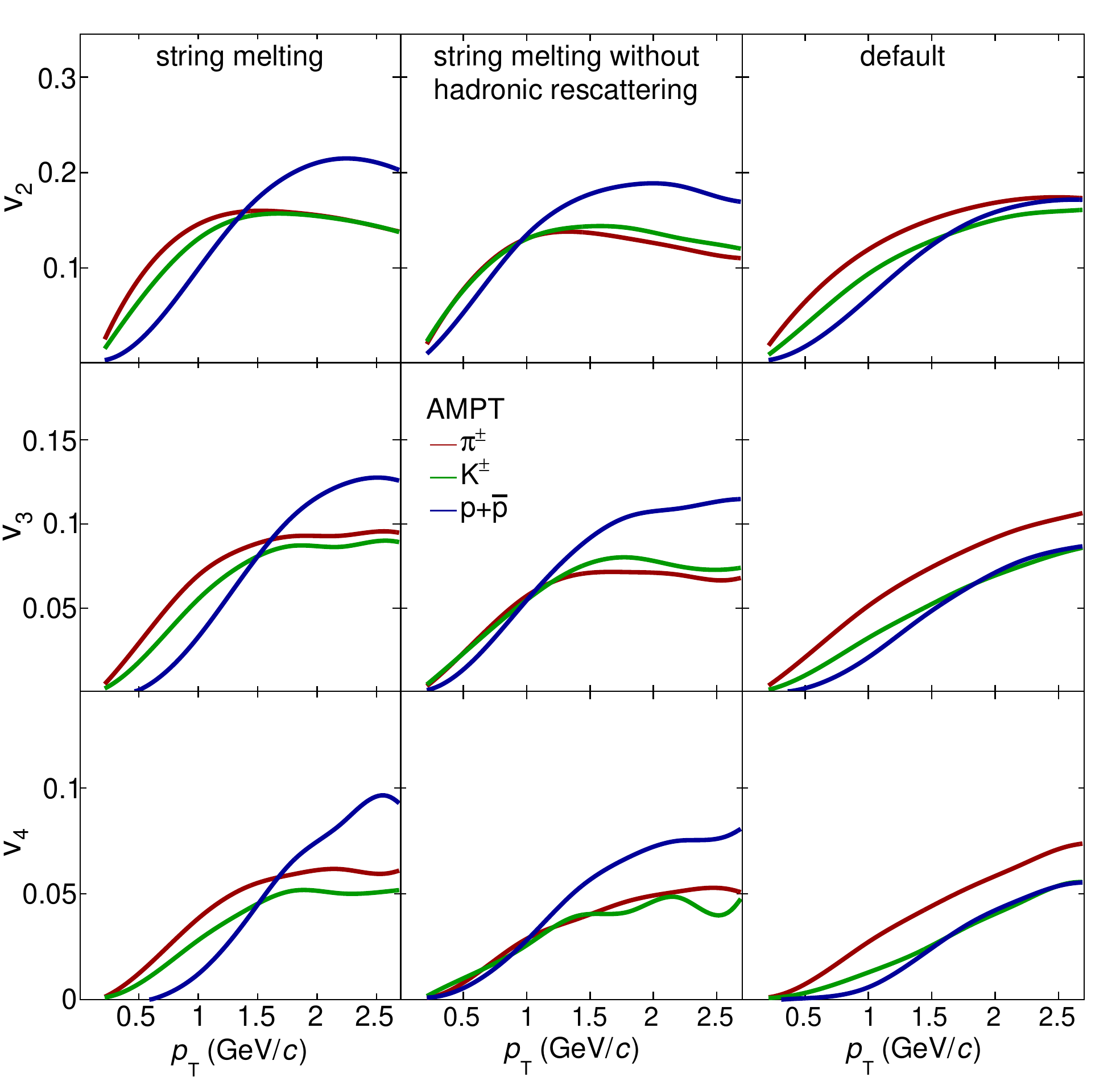}
\end{center}
\captionof{figure}{The \vtwoAA(\pT), \vthreeAA(\pT) and \vfourAA(\pT) in 20--30\% central Pb--Pb collisions at \sNN, obtained using the string melting, with (left) and without (middle) hadronic rescattering, and the default (right) versions.}
\label{AMPTconfig_2030}
\end{figure}

\subsubsection{Comparison with AMPT}
\label{Section:AMPT}

\begin{figure}[!htb]
\begin{center}
\includegraphics[scale=0.55]{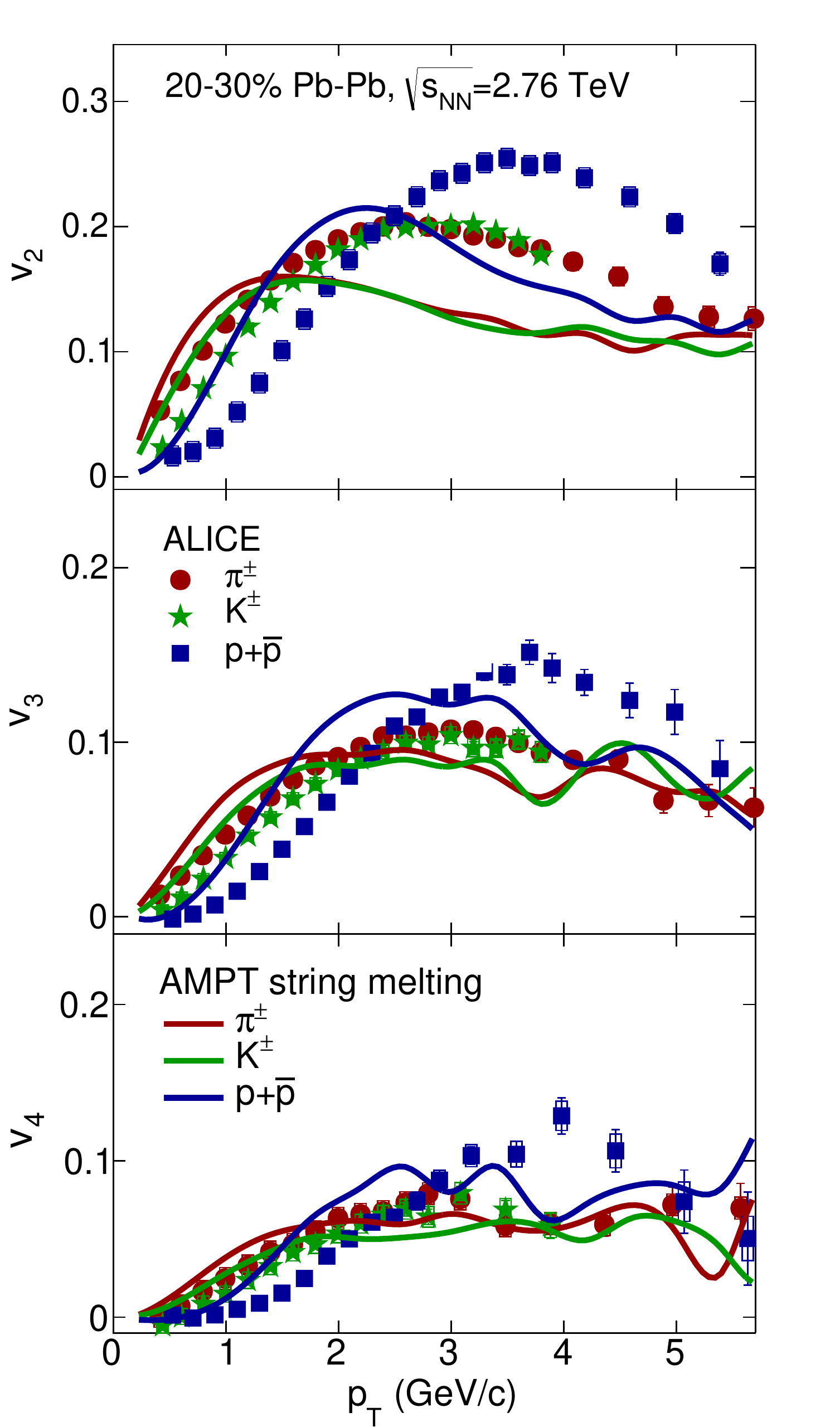}
\end{center}
\captionof{figure}{The \vtwo(\pT), \vthree(\pT) and \vfour(\pT) for \pion, \kaon~and \proton~measured in Pb--Pb collisions at \sNN~compared to AMPT (with the string melting option) in the 20--30\% centrality range.}
\label{AMPTvsALICE_2030}
\end{figure}

In addition to the hydrodynamical calculations discussed in the previous paragraphs, three different versions of AMPT~\cite{Zhang:1999bd,Lin:2000cx,Lin:2004en} are studied in this article. The AMPT model can be run in two main configurations: the default and the string melting. In the default version, partons are recombined with the parent strings when they stop interacting. The resulting strings are later converted into hadrons using the Lund string fragmentation model~\cite{Andersson:1986gw,NilssonAlmqvist:1986rx}. In the string melting version, the initial strings are melted into partons whose interactions are described by a parton cascade model~\cite{Zhang:1997ej}. These partons are then combined into the final-state hadrons via a quark coalescence model. In both configurations a final-state hadronic rescattering is implemented which also includes resonance decays. The third version presented in this article is based on the string melting configuration, in which the hadronic rescattering phase is switched off to study its influence to the development of anisotropic flow. The input parameters used in all cases are: $\alpha_s = 0.33$, a partonic cross-section of 1.5~mb, while the Lund string fragmentation parameters were set to $\alpha = 0.5$ and $b = 0.9$~GeV$^{-2}$. 

Figure~\ref{AMPTconfig_2030} presents the $p_{\mathrm{T}}$-differential $v_2$ (first row), $v_3$ (middle row) and $v_4$ (bottom row) for pions, kaons and protons for the 20--30\% centrality interval. Each column presents the results of one of the three AMPT versions discussed above. The string melting AMPT version (left column) predicts a distinct mass ordering at low values of transverse momentum as well as a lower value of $v_n$ for mesons compared to baryons in the intermediate \pT~region for all harmonics, similar to what is observed in the experimental measurements. On the other hand, the version with string melting but without the hadronic rescattering contribution (middle column) can only reproduce the particle type grouping at intermediate \pT~values.  Finally, the default AMPT version is only able to reproduce the mass ordering in the low \pT~region. These observations suggest that the string melting and the final-state hadronic rescattering are responsible for the particle type grouping at intermediate \pT~and the mass ordering at low \pT, respectively.

Since the AMPT string melting version is able to reproduce the main features of the experimental measurement throughout the reported \pT~range, the corresponding results are compared with the data points in Fig.~\ref{AMPTvsALICE_2030}. It is seen that although this version of AMPT reproduces both the mass ordering and the particle type grouping at low and intermediate \pT~for all harmonics, it fails to quantitatively reproduce the measurements. In order to understand the origin of this discrepancy both spectra and the \pT-differential \vtwo~for different particle species for the 20--30\% centrality interval in AMPT were fitted with a blast-wave parametrisation~\cite{Retiere:2003kf}. The results were compared to the analogous parameters obtained from the experiment~\cite{Abelev:2013vea}. It turns out that the radial flow in AMPT is around 25\% lower than the measured value at the LHC. As the radial flow is essential in shaping the \pT~dependence of \vn~we suppose that the unrealistically low radial flow in AMPT is responsible for the quantitative disagreement.\\

\section{Conclusions}
\label{Sec:Conclusions}

In this article, a measurement of non-flow subtracted flow harmonics, \vtwo, \vthree, \vfour~and \vfive~as a function of transverse momentum for \pion, \kaon~and \proton~for different centrality intervals (0--1\% up to 40--50\%) in Pb--Pb collisions at \sNN~are reported. The \vn~coefficients are calculated with the Scalar Product method, selecting the identified hadron under study and the reference flow particles from different, non-overlapping pseudorapidity regions. Correlations not related to the common symmetry planes (i.e.~non-flow) were estimated based on pp collisions and were subtracted from the measurements.

The validity of this subtraction procedure was checked by repeating the analysis using different charge combinations (i.e.~positive--positive and negative--negative) for the identified hadrons and the reference particles collisions as well as applying different pseudorapidity gaps ($\Delta\eta$) between them, in both Pb--Pb and pp collisions. The results after the subtraction in both cases did not exhibit any systematic change in \vn(\pT) with respect to the default ones for any particle species or centrality.

All flow harmonics exhibit an increase in peripheral compared to central collisions. This increase is more pronounced for \vtwo~than for the higher harmonics. This indicates that \vtwo~reflects mainly the geometry of the system, while higher order flow harmonics are primarily generated by event-by-event fluctuations of the initial energy density profile. This is also supported by the observation of a significant non-zero value of \vn$>0$~in ultra-central (i.e.~0--1$\%$) collisions. In this centrality interval of Pb--Pb collisions, \vthree~and \vfour~become gradually larger than \vtwo~at a transverse momentum value which increases with increasing order of the flow harmonics and particle mass. In addition, a distinct mass ordering is observed for all \vn~coefficients in all centrality intervals in the low transverse momentum region, i.e.~for $p_{\mathrm{T}} < 3$ GeV/$c$. Furthermore, the \vn(\pT) values show a crossing between \pion, \kaon~and \proton, that takes place at different \pT~values depending on the centrality and the order of the flow harmonic. These observations are attributed to the interplay between not only \vtwo~but also the higher flow harmonics and radial flow. For transverse momentum values beyond the crossing point between different particle species (i.e.~for $p_{\mathrm{T}} > 3$ GeV/$c$), the values of \vn~for baryons are larger than for mesons. The NCQ scaling holds for \vtwo~at an approximate level of $\pm$20\% which is in agreement with~\cite{Abelev:2014pua}. For higher harmonics this scaling holds at a similar level within the current level of statistical and systematic uncertainties.

In the low momentum region, hydrodynamic calculations based on iEBE-VISHNU describe \vtwo~for all three particle species and \vthree~and \vfour~of pions fairly well. For kaons and protons the model seems to overpredict \vthree~and \vfour~in almost all centrality intervals. Finally, the comparison of different AMPT configurations with the measurements highlights the importance of the final state hadronic rescattering stage and of particle production via the coalescence mechanism to the development of the mass ordering and the particle type grouping at low and intermediate transverse momentum values, respectively. The AMPT with string melting is able to describe qualitatively both of these features that the experimental data exhibit. However, it fails to quantitatively describe the measurements, probably due to a significantly smaller radial flow compared to the experiment.

%
%

\newenvironment{acknowledgement}{\relax}{\relax}
\begin{acknowledgement}
\section*{Acknowledgements}
The ALICE Collaboration would like to thank all its engineers and technicians for their invaluable contributions to the construction of the experiment and the CERN accelerator teams for the outstanding performance of the LHC complex.
The ALICE Collaboration gratefully acknowledges the resources and support provided by all Grid centres and the Worldwide LHC Computing Grid (WLCG) collaboration.
The ALICE Collaboration acknowledges the following funding agencies for their support in building and
running the ALICE detector:
State Committee of Science,  World Federation of Scientists (WFS)
and Swiss Fonds Kidagan, Armenia,
Conselho Nacional de Desenvolvimento Cient\'{\i}fico e Tecnol\'{o}gico (CNPq), Financiadora de Estudos e Projetos (FINEP),
Funda\c{c}\~{a}o de Amparo \`{a} Pesquisa do Estado de S\~{a}o Paulo (FAPESP);
National Natural Science Foundation of China (NSFC), the Chinese Ministry of Education (CMOE)
and the Ministry of Science and Technology of China (MSTC);
Ministry of Education and Youth of the Czech Republic;
Danish Natural Science Research Council, the Carlsberg Foundation and the Danish National Research Foundation;
The European Research Council under the European Community's Seventh Framework Programme;
Helsinki Institute of Physics and the Academy of Finland;
French CNRS-IN2P3, the `Region Pays de Loire', `Region Alsace', `Region Auvergne' and CEA, France;
German BMBF and the Helmholtz Association;
General Secretariat for Research and Technology, Ministry of
Development, Greece;
Hungarian OTKA and National Office for Research and Technology (NKTH);
Department of Atomic Energy and Department of Science and Technology of the Government of India;
Istituto Nazionale di Fisica Nucleare (INFN) and Centro Fermi -
Museo Storico della Fisica e Centro Studi e Ricerche "Enrico
Fermi", Italy;
MEXT Grant-in-Aid for Specially Promoted Research, Ja\-pan;
Joint Institute for Nuclear Research, Dubna;
National Research Foundation of Korea (NRF);
CONACYT, DGAPA, M\'{e}xico, ALFA-EC and the EPLANET Program
(European Particle Physics Latin American Network)
Stichting voor Fundamenteel Onderzoek der Materie (FOM) and the Nederlandse Organisatie voor Wetenschappelijk Onderzoek (NWO), Netherlands;
Research Council of Norway (NFR);
Polish Ministry of Science and Higher Education;
National Science Centre, Poland;
 Ministry of National Education/Institute for Atomic Physics and CNCS-UEFISCDI - Romania;
Ministry of Education and Science of Russian Federation, Russian
Academy of Sciences, Russian Federal Agency of Atomic Energy,
Russian Federal Agency for Science and Innovations and The Russian
Foundation for Basic Research;
Ministry of Education of Slovakia;
Department of Science and Technology, South Africa;
CIEMAT, EELA, Ministerio de Econom\'{i}a y Competitividad (MINECO) of Spain, Xunta de Galicia (Conseller\'{\i}a de Educaci\'{o}n),
CEA\-DEN, Cubaenerg\'{\i}a, Cuba, and IAEA (International Atomic Energy Agency);
Swedish Research Council (VR) and Knut $\&$ Alice Wallenberg
Foundation (KAW);
Ukraine Ministry of Education and Science;
United Kingdom Science and Technology Facilities Council (STFC);
The United States Department of Energy, the United States National
Science Foundation, the State of Texas, and the State of Ohio.    
\end{acknowledgement}

\bibliographystyle{utphys}   
\bibliography{paperCDS}


\appendix
\clearpage
\section{Additional figures}
\label{Sec:Appendix}

\subsection{Integrated \rawvn}
\label{Subsec:IntegratedVn}

\begin{figure}[htb]
\begin{center}
\includegraphics[scale=0.8]{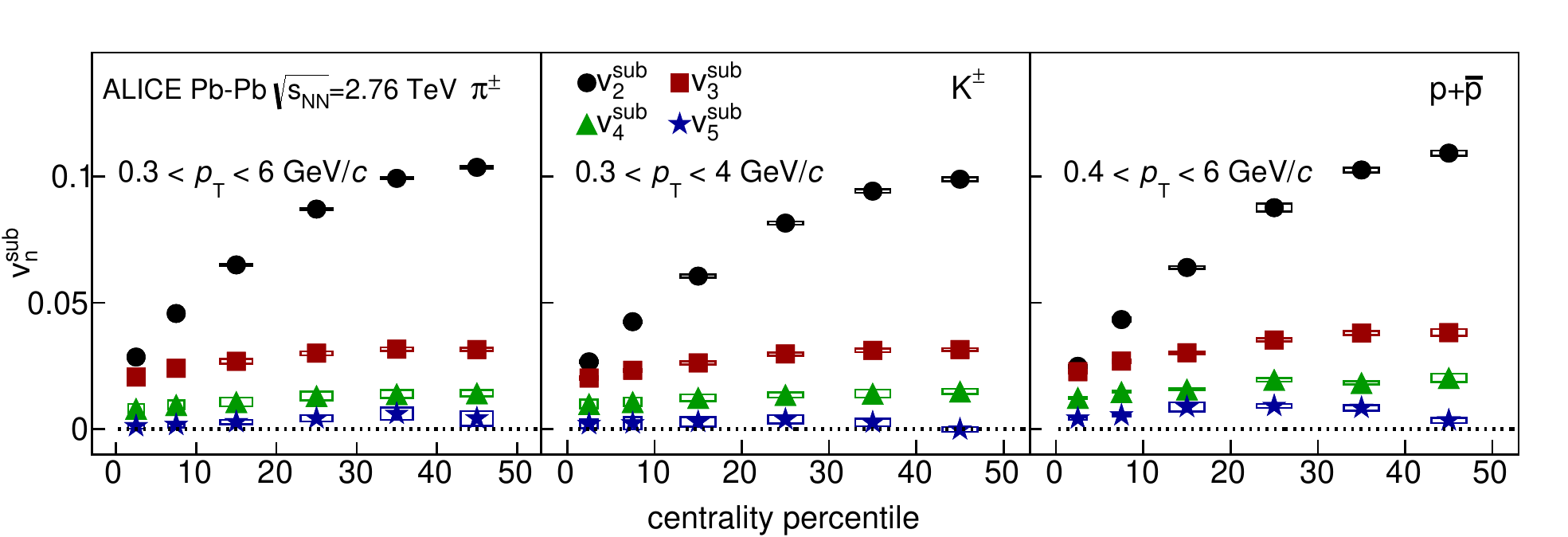}
\end{center}
\caption{The \rawvtwo, \rawvthree~and \rawvfour~integrated over the \pT~range $0.3<$\pT$<6$ \GeV~for \pion~(left), $0.3<$\pT$<4$ \GeV~for \kaon~(middle) and $0.4<$\pT$<6$ \GeV~for \proton~(right) as a function of centrality intervals in Pb--Pb collisions at \sNN.}
\label{integratedvn}
\end{figure}

\subsection{NCQ scaling}
\label{Subsec:NCQScaling}

\begin{figure}[htb]
\begin{center}
\includegraphics[width=0.49\textwidth]{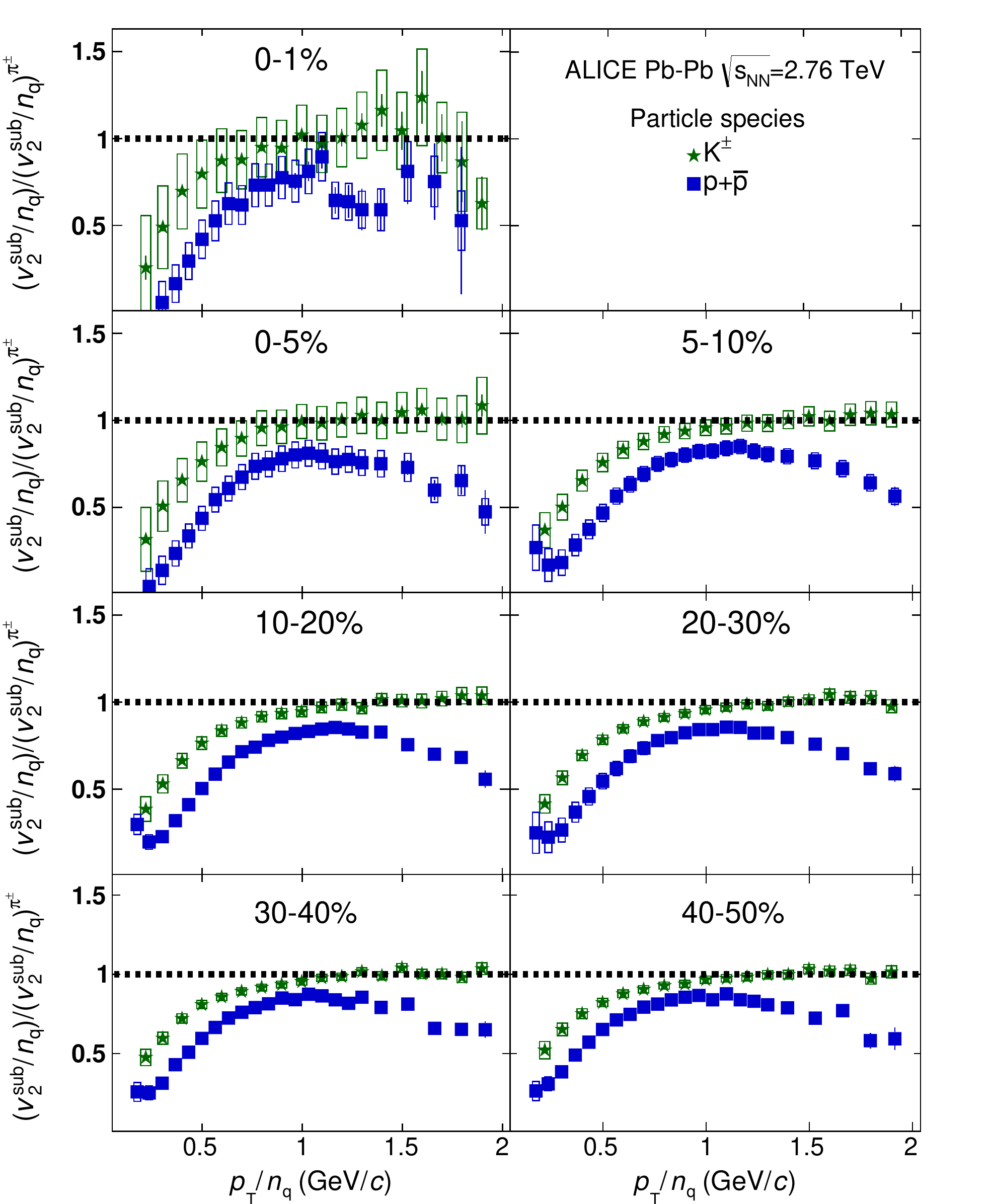}
\includegraphics[width=0.49\textwidth]{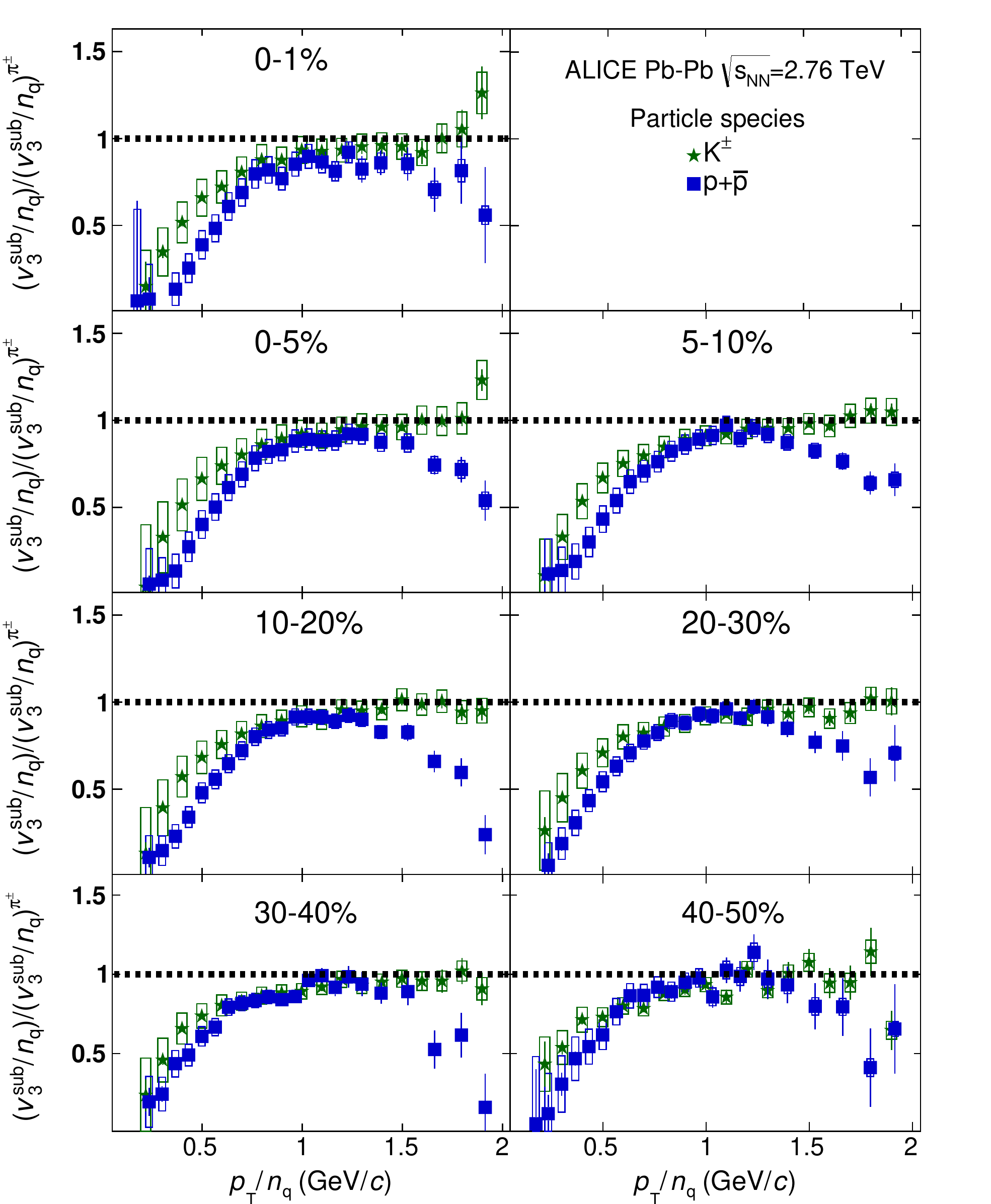}
\end{center}
\caption{Left: the \pTnq~dependence of the double ratio of \vtwonq~for \kaon~and \proton~relative to a fit to \vtwonq~of \pion~for Pb--Pb collisions in various centrality intervals at \sNN. Right: the same for \vthreenq.}
\label{v2Ratio_NCQ}
\end{figure}
\begin{figure}[htb]
\begin{center}
\includegraphics[width=0.49\textwidth]{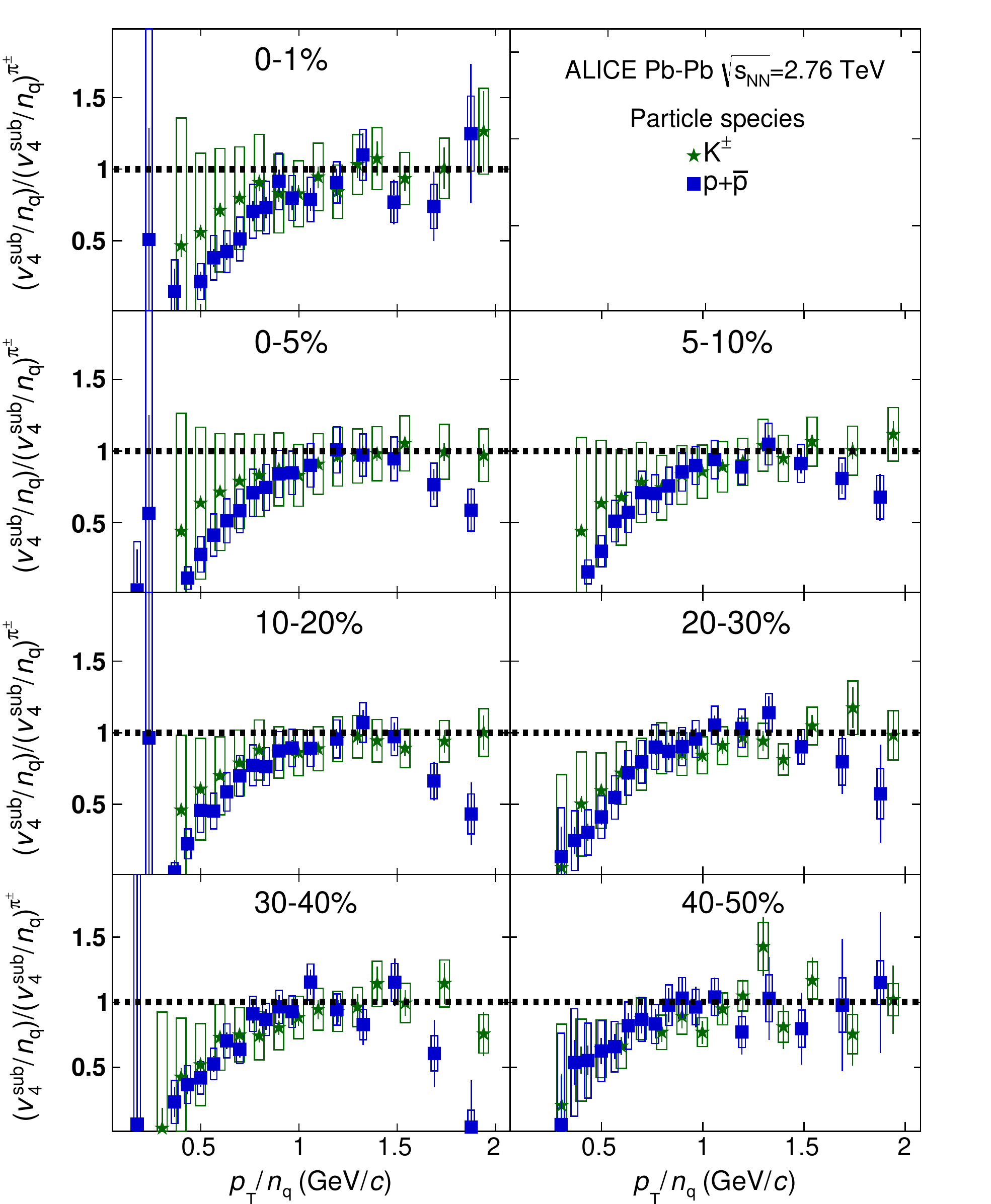}
\includegraphics[width=0.49\textwidth]{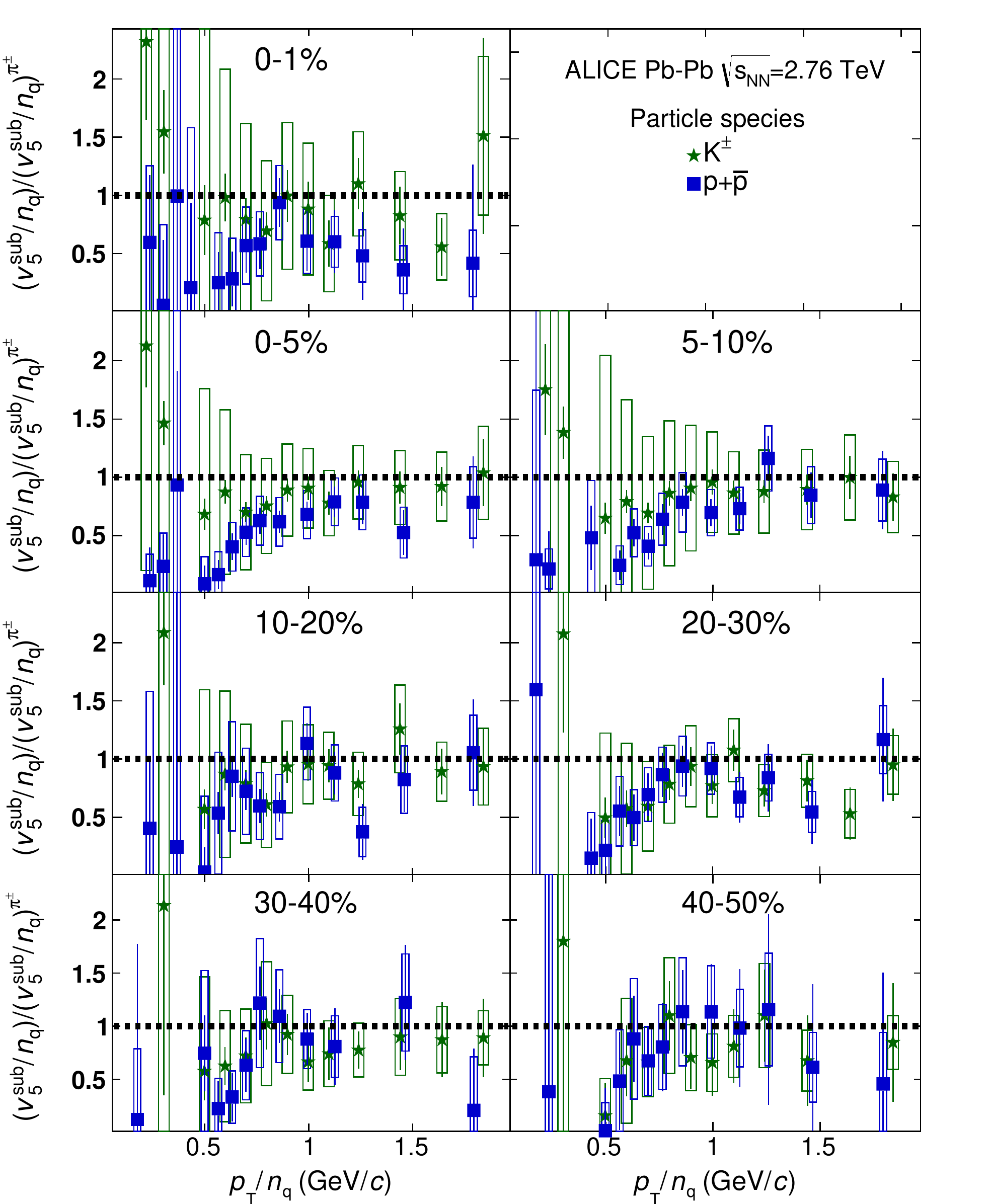}
\end{center}
\caption{Left: the \pTnq~dependence of the double ratio of \vfournq~for \kaon~and \proton~relative to a fit to \vfournq~of \pion~for Pb--Pb collisions in various centrality intervals at \sNN. Right: the same for \vfivenq.}
\label{v4Ratio_NCQ}
\end{figure}


\clearpage
\subsection{$KE_{\mathrm{T}}$ scaling}
\label{Subsec:KETScaling}
It was suggested at RHIC to extend the scaling to lower \pT~values by studying the transverse kinetic energy dependence of anisotropic flow harmonics. Transverse kinetic energy is defined as $KE_{\mathrm{T}} = m_{\mathrm{T}}-m_{0}$, where $m_{\mathrm{T}}= \sqrt{m_{0}^{2}+p_{\mathrm{T}}^{2}}$ is the transverse mass. 

\begin{figure}[htb]
\begin{center}
\includegraphics[width=0.49\textwidth]{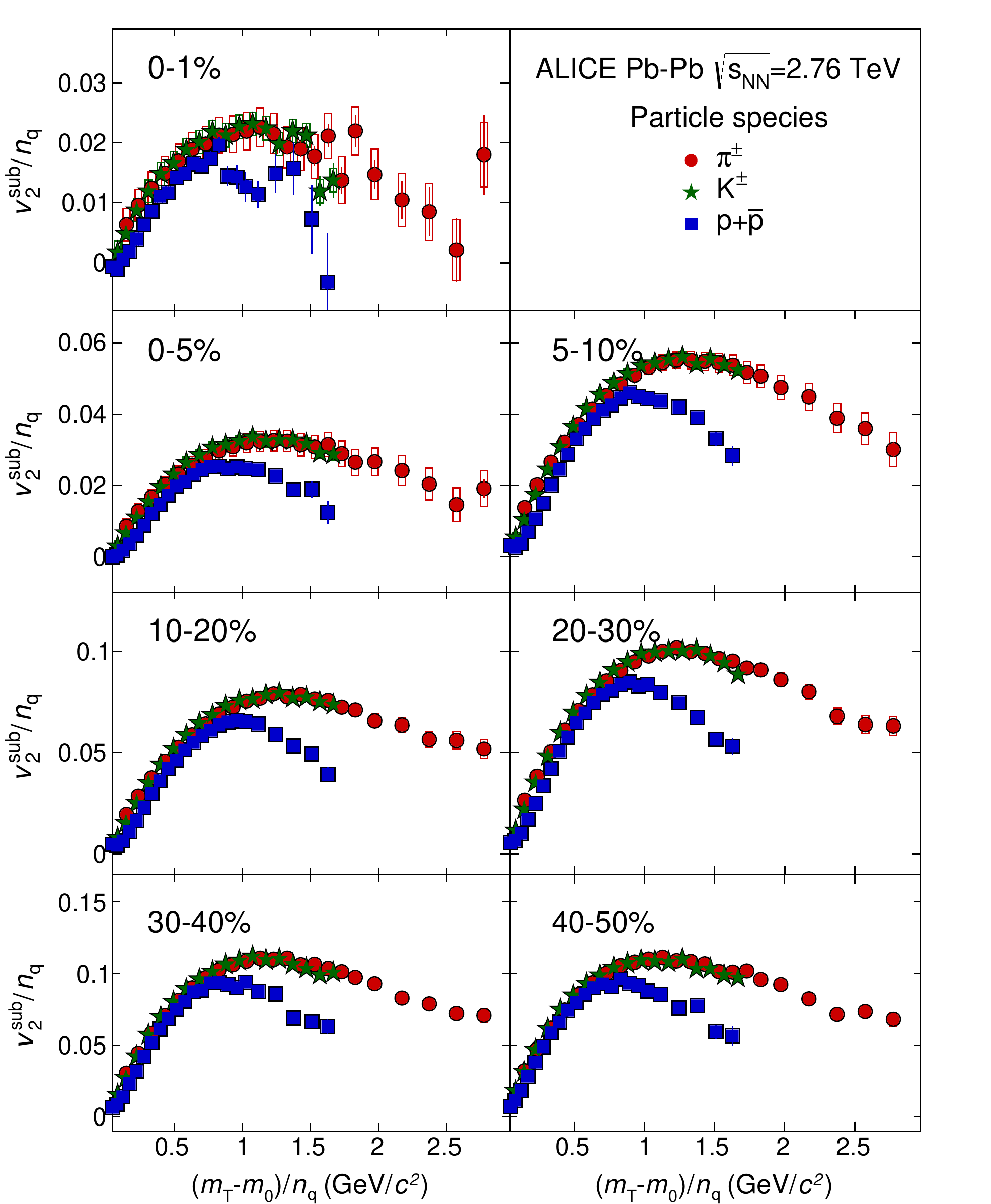}
\includegraphics[width=0.49\textwidth]{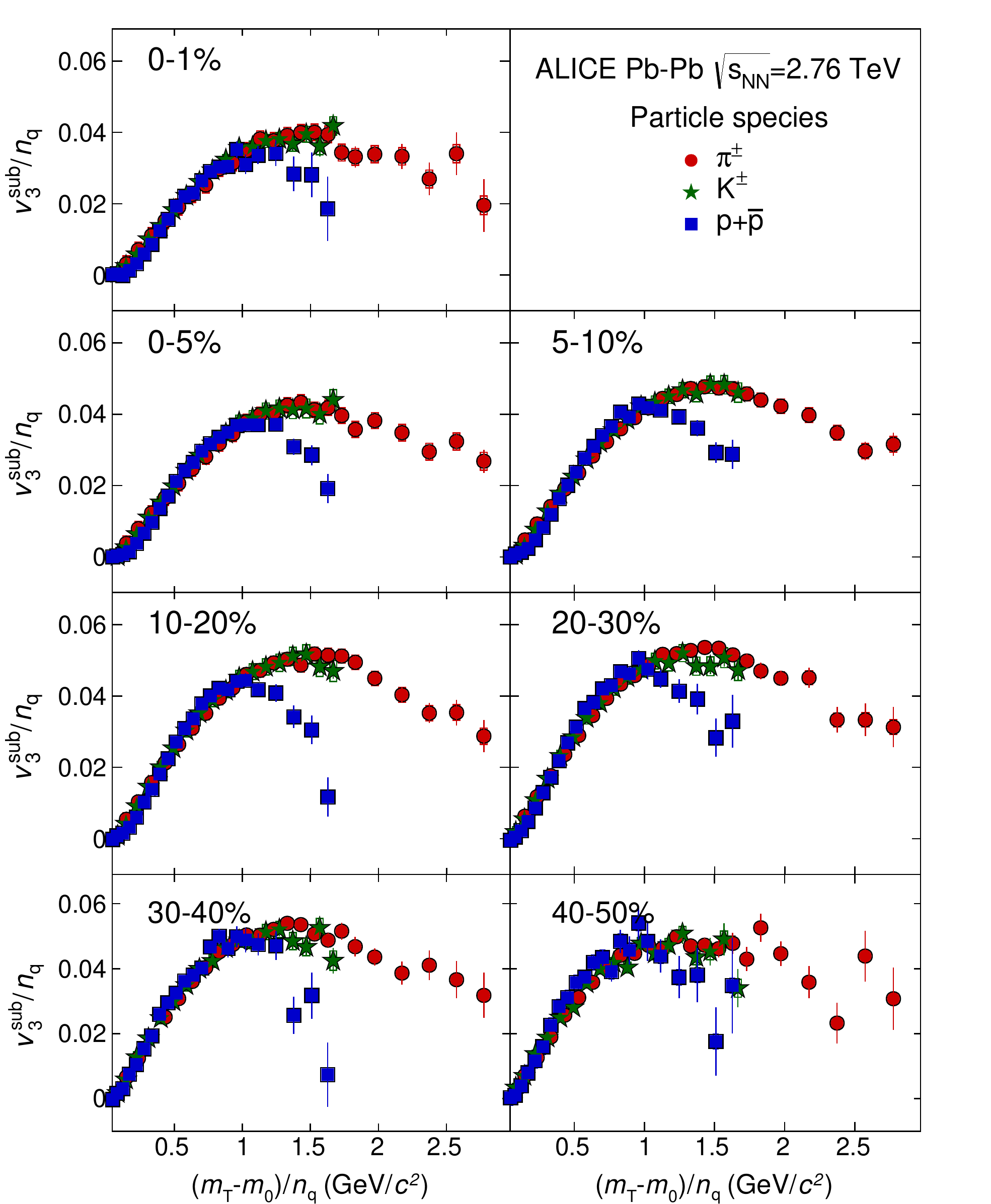}
\end{center}
\caption{The \pTmT-dependence of \vtwonq~(left) and \vthreenq~(right) for \pion, \kaon~and \proton~for Pb-Pb collisions in various centrality intervals at \sNN.}
\label{v2_KET}
\end{figure}

\begin{figure}
\begin{center}
\includegraphics[width=0.49\textwidth]{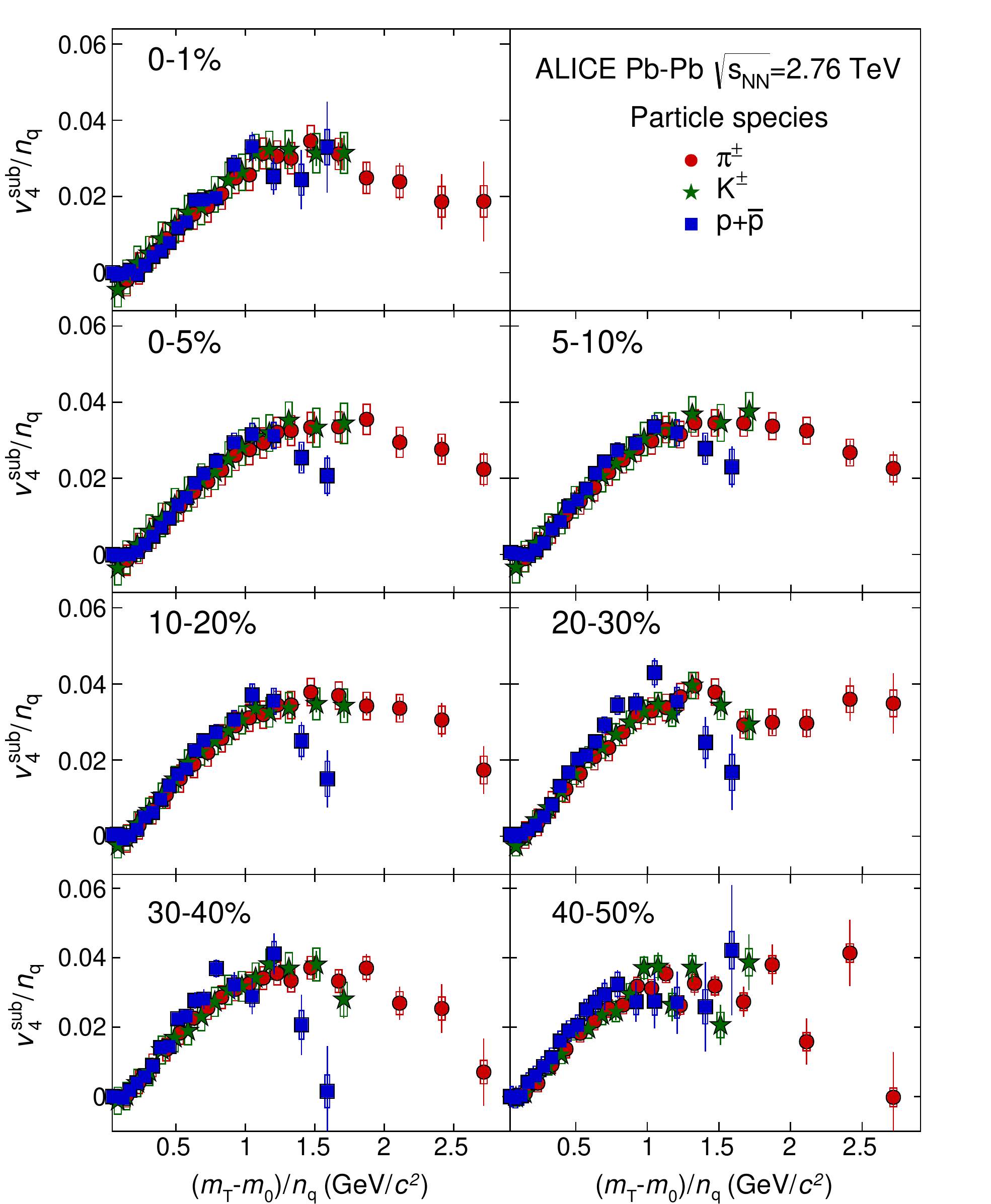}
\includegraphics[width=0.49\textwidth]{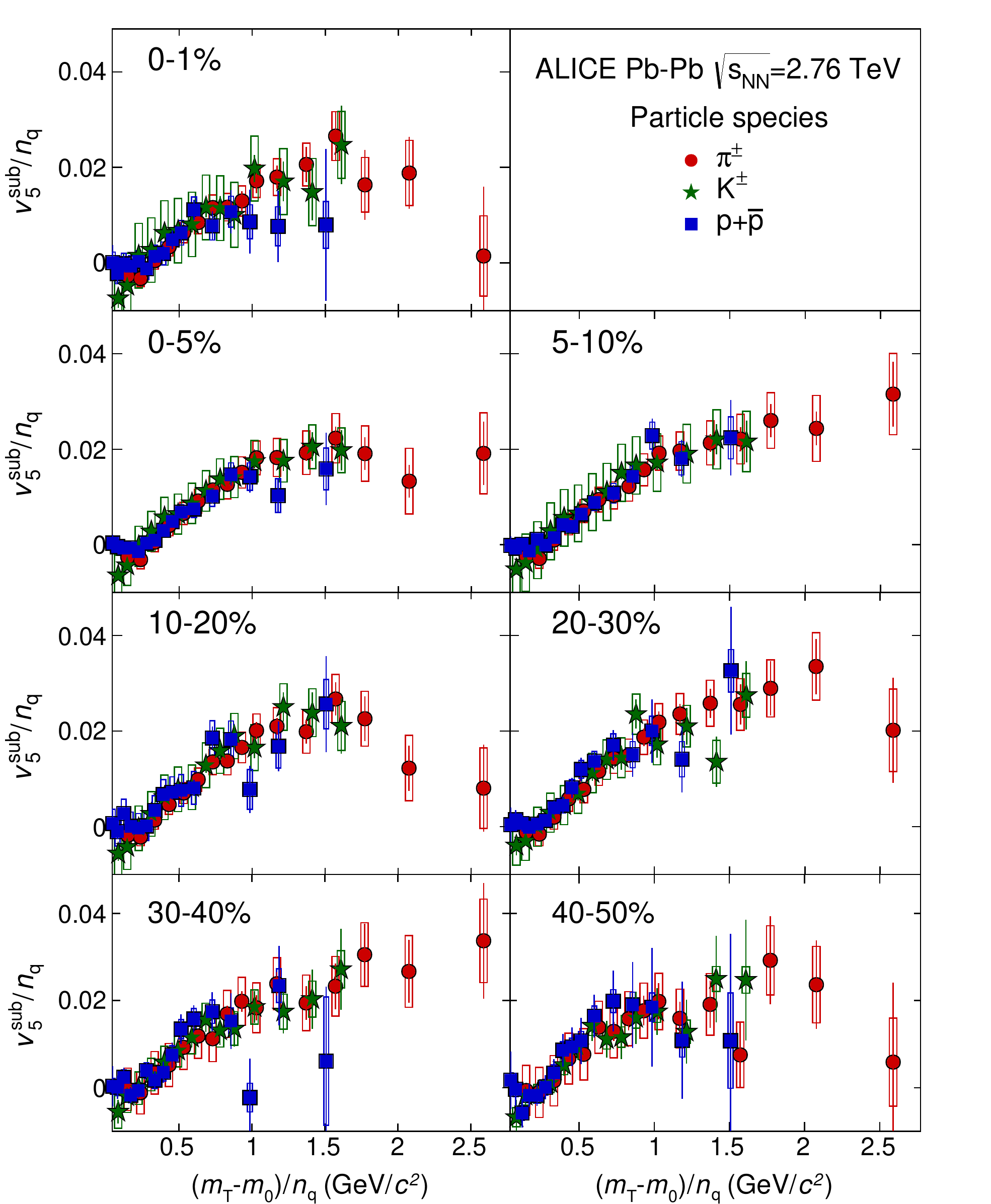}
\end{center}
\caption{The \pTmT-dependence of \vfournq~(left) and \vfivenq~(right) for \pion, \kaon~and \proton~for Pb-Pb collisions in various centrality intervals at \sNN.}
\label{v4_KET}
\end{figure}

\begin{figure}[htb]
\begin{center}
\includegraphics[width=0.49\textwidth]{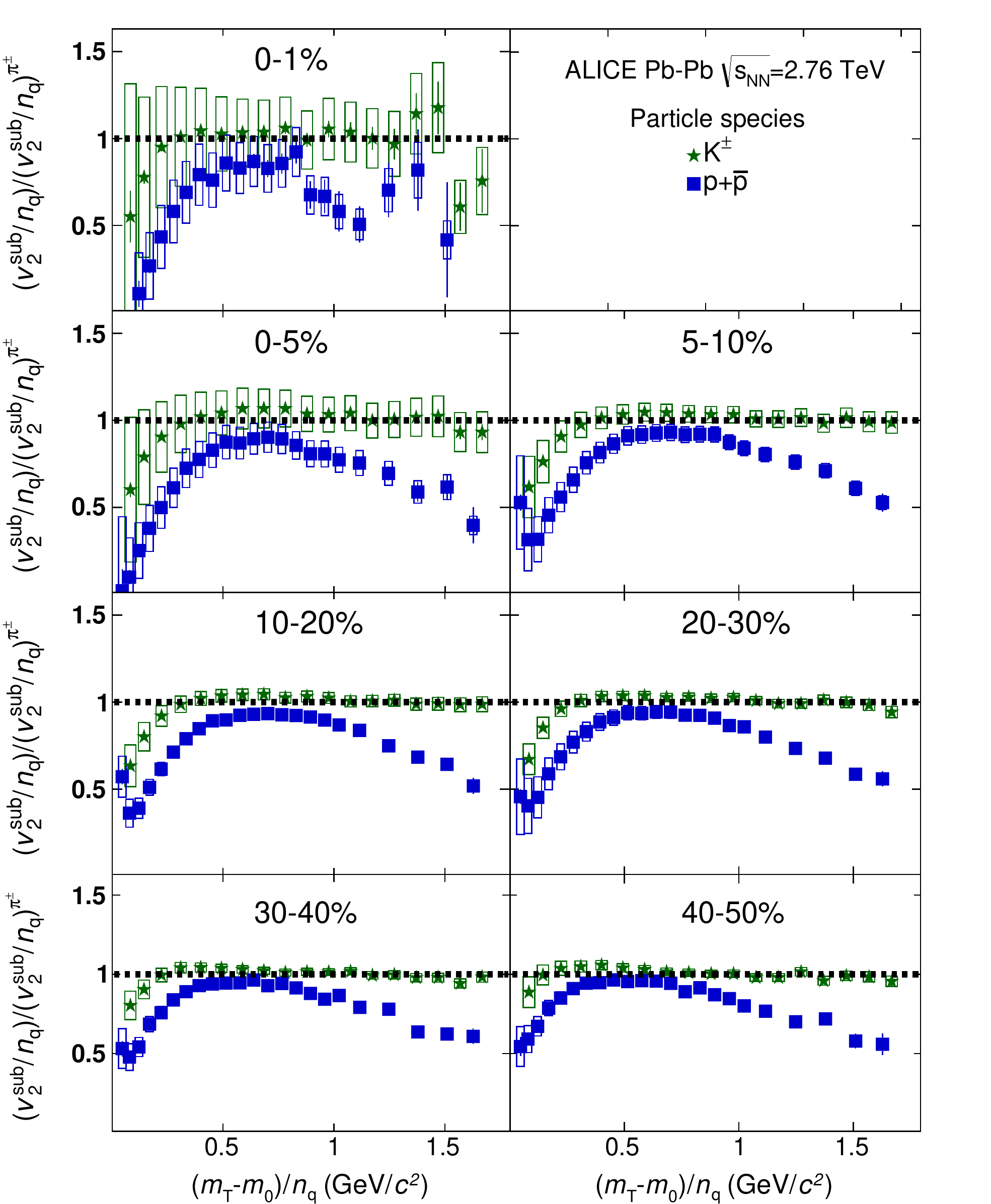}
\includegraphics[width=0.49\textwidth]{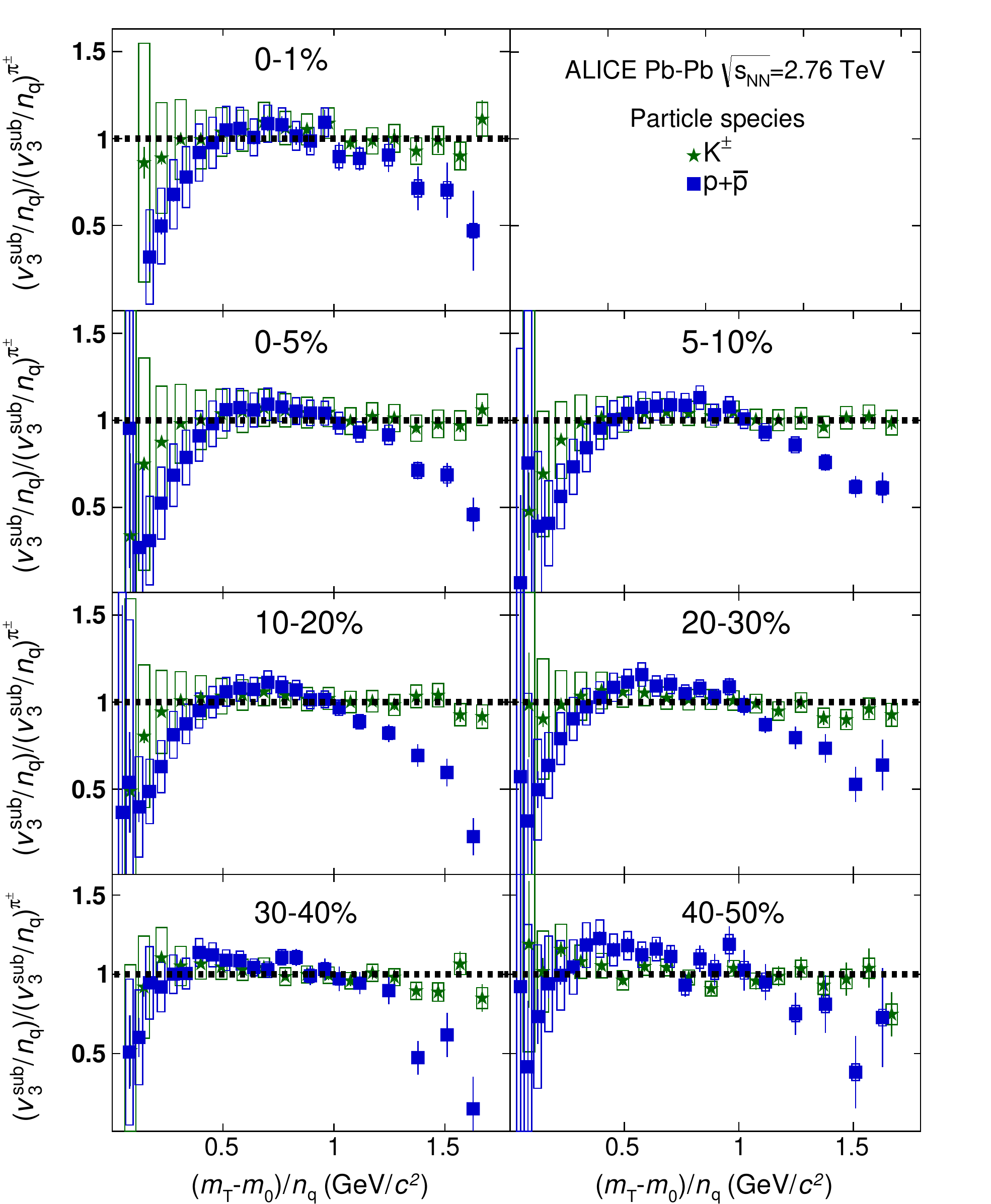}
\end{center}
\caption{Left: the \pTmT~dependence of the double ratio of \vtwonq~for \kaon~and \proton~relative to a fit to \vtwonq~of \pion~for Pb--Pb collisions in various centrality intervals at \sNN. Right: the same for \vthreenq.}
\label{v2Ratio_KET}
\end{figure}

\begin{figure}[htb]
\begin{center}
\includegraphics[width=0.49\textwidth]{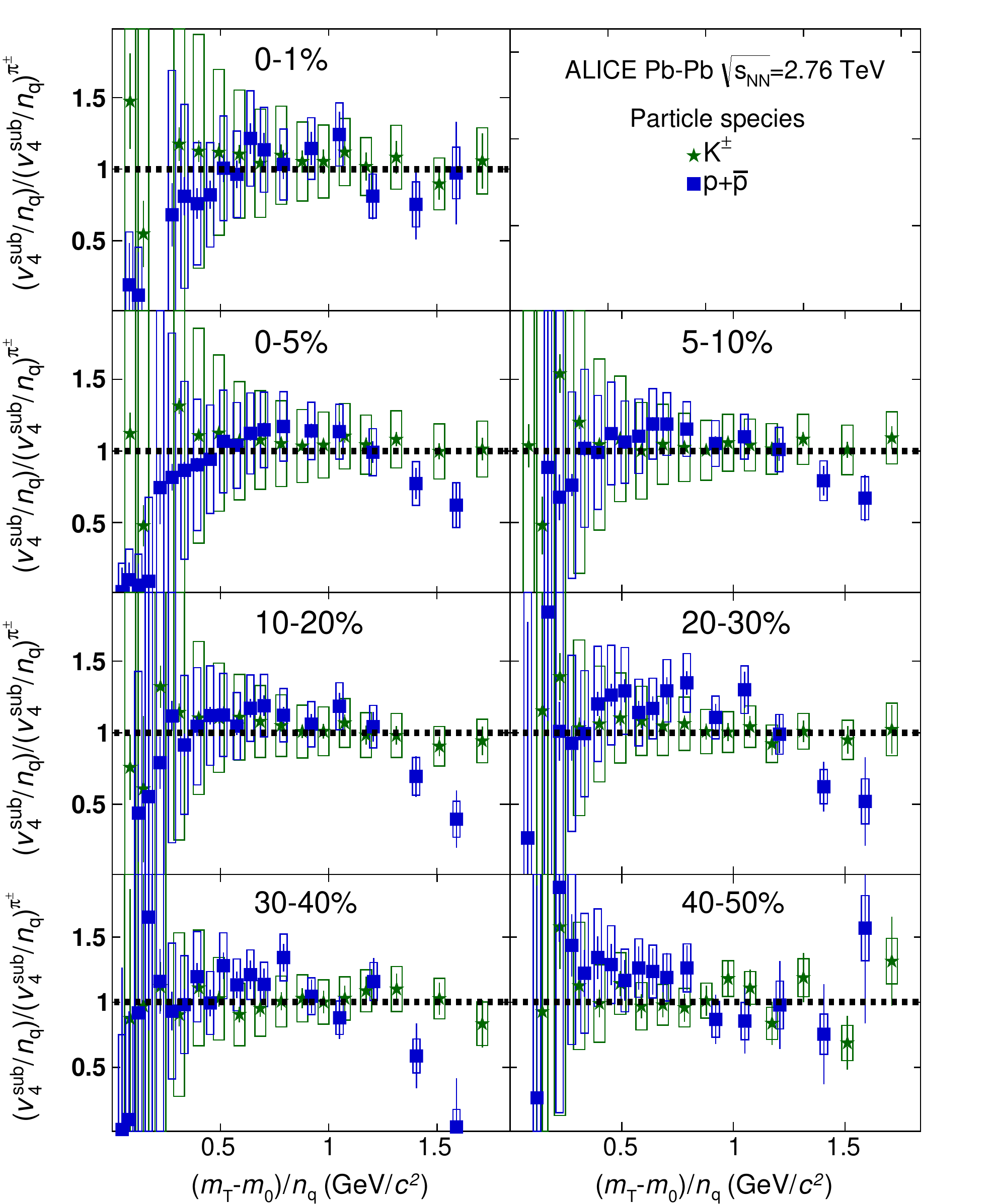}
\includegraphics[width=0.49\textwidth]{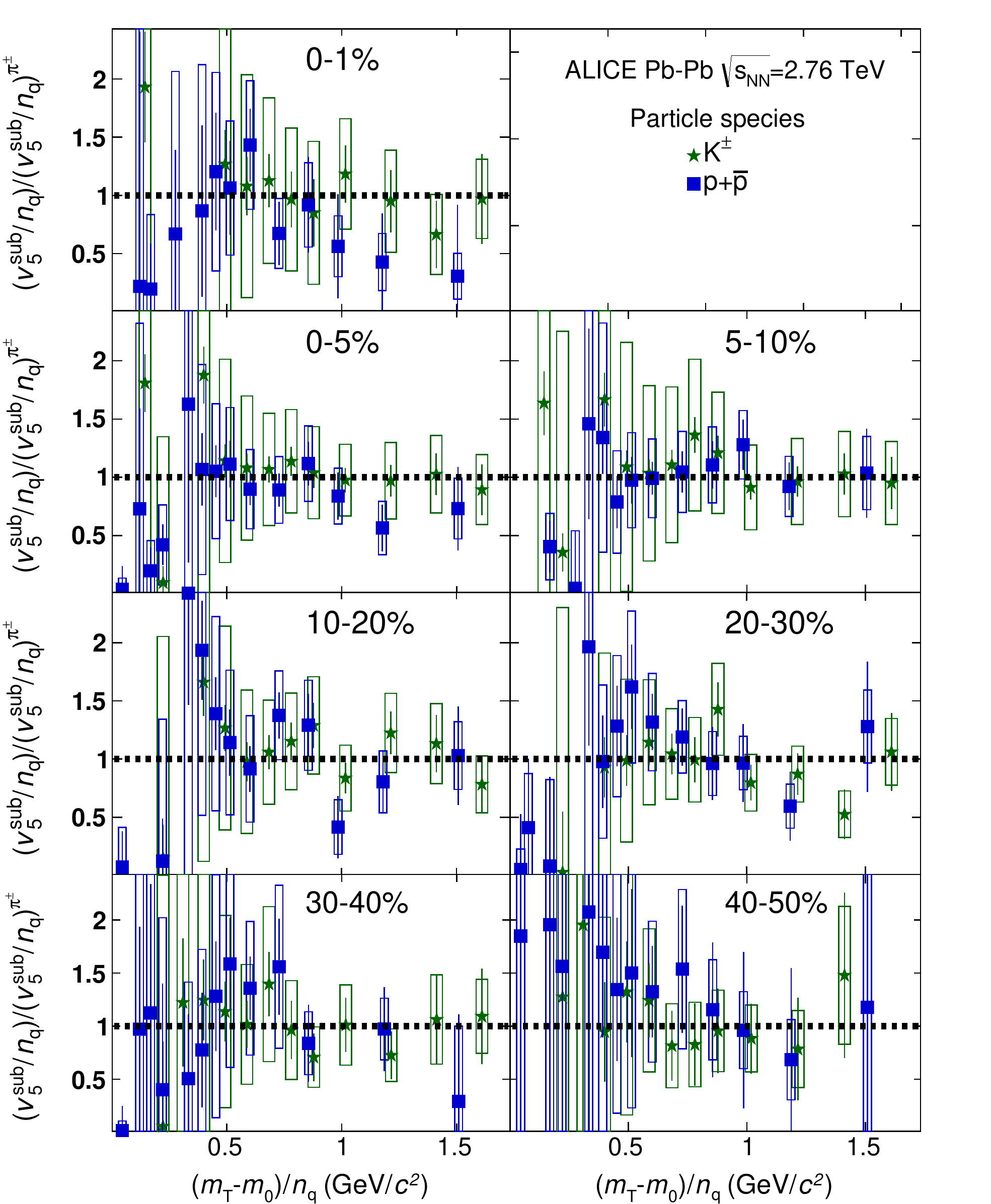}
\end{center}
\caption{Left: the \pTmT~dependence of the double ratio of \vfournq~for \kaon~and \proton~relative to a fit to \vfournq~of \pion~for Pb--Pb collisions in various centrality intervals at \sNN. Right: the same for \vfivenq.}
\label{v4Ratio_KET}
\end{figure}

\clearpage
\newpage
\section{The ALICE Collaboration}
\label{app:collab}



\begingroup
\small
\begin{flushleft}
J.~Adam$^\textrm{\scriptsize 39}$,
D.~Adamov\'{a}$^\textrm{\scriptsize 85}$,
M.M.~Aggarwal$^\textrm{\scriptsize 89}$,
G.~Aglieri Rinella$^\textrm{\scriptsize 35}$,
M.~Agnello$^\textrm{\scriptsize 112}$\textsuperscript{,}$^\textrm{\scriptsize 31}$,
N.~Agrawal$^\textrm{\scriptsize 48}$,
Z.~Ahammed$^\textrm{\scriptsize 136}$,
S.~Ahmad$^\textrm{\scriptsize 18}$,
S.U.~Ahn$^\textrm{\scriptsize 69}$,
S.~Aiola$^\textrm{\scriptsize 140}$
% ...
\renewcommand\labelenumi{\textsuperscript{\theenumi}~}

\section*{Affiliation notes}
\renewcommand\theenumi{\roman{enumi}}
\begin{Authlist}
\item \Adef{0}Deceased
\item \Adef{idp1832880}{Also at: Georgia State University, Atlanta, Georgia, United States}
\item \Adef{idp3268336}{Also at: Also at Department of Applied Physics, Aligarh Muslim University, Aligarh, India}
\item \Adef{idp4000576}{Also at: M.V. Lomonosov Moscow State University, D.V. Skobeltsyn Institute of Nuclear, Physics, Moscow, Russia}
\end{Authlist}

\section*{Collaboration Institutes}
\renewcommand\theenumi{\arabic{enumi}~}

$^{1}$A.I. Alikhanyan National Science Laboratory (Yerevan Physics Institute) Foundation, Yerevan, Armenia
\\
$^{2}$Benem\'{e}rita Universidad Aut\'{o}noma de Puebla, Puebla, Mexico
\\
$^{3}$Bogolyubov Institute for Theoretical Physics, Kiev, Ukraine
\\
$^{4}$Bose Institute, Department of Physics 
and Centre for Astroparticle Physics and Space Science (CAPSS), Kolkata, India
\\
$^{5}$Budker Institute for Nuclear Physics, Novosibirsk, Russia
\\
$^{6}$California Polytechnic State University, San Luis Obispo, California, United States
\\
$^{7}$Central China Normal University, Wuhan, China
\\
$^{8}$Centre de Calcul de l'IN2P3, Villeurbanne, Lyon, France
\\
$^{9}$Centro de Aplicaciones Tecnol\'{o}gicas y Desarrollo Nuclear (CEADEN), Havana, Cuba
\\
$^{10}$Centro de Investigaciones Energ\'{e}ticas Medioambientales y Tecnol\'{o}gicas (CIEMAT), Madrid, Spain
\\
$^{11}$Centro de Investigaci\'{o}n y de Estudios Avanzados (CINVESTAV), Mexico City and M\'{e}rida, Mexico
\\
$^{12}$Centro Fermi - Museo Storico della Fisica e Centro Studi e Ricerche ``Enrico Fermi', Rome, Italy
\\
$^{13}$Chicago State University, Chicago, Illinois, United States
\\
$^{14}$China Institute of Atomic Energy, Beijing, China
\\
$^{15}$Commissariat \`{a} l'Energie Atomique, IRFU, Saclay, France
\\
$^{16}$COMSATS Institute of Information Technology (CIIT), Islamabad, Pakistan
\\
$^{17}$Departamento de F\'{\i}sica de Part\'{\i}culas and IGFAE, Universidad de Santiago de Compostela, Santiago de Compostela, Spain
\\
$^{18}$Department of Physics, Aligarh Muslim University, Aligarh, India
\\
$^{19}$Department of Physics, Ohio State University, Columbus, Ohio, United States
\\
$^{20}$Department of Physics, Sejong University, Seoul, South Korea
\\
$^{21}$Department of Physics, University of Oslo, Oslo, Norway
\\
$^{22}$Department of Physics and Technology, University of Bergen, Bergen, Norway
\\
$^{23}$Dipartimento di Fisica dell'Universit\`{a} 'La Sapienza'
and Sezione INFN, Rome, Italy
\\
$^{24}$Dipartimento di Fisica dell'Universit\`{a}
and Sezione INFN, Cagliari, Italy
\\
$^{25}$Dipartimento di Fisica dell'Universit\`{a}
and Sezione INFN, Trieste, Italy
\\
$^{26}$Dipartimento di Fisica dell'Universit\`{a}
and Sezione INFN, Turin, Italy
\\
$^{27}$Dipartimento di Fisica e Astronomia dell'Universit\`{a}
and Sezione INFN, Bologna, Italy
\\
$^{28}$Dipartimento di Fisica e Astronomia dell'Universit\`{a}
and Sezione INFN, Catania, Italy
\\
$^{29}$Dipartimento di Fisica e Astronomia dell'Universit\`{a}
and Sezione INFN, Padova, Italy
\\
$^{30}$Dipartimento di Fisica `E.R.~Caianiello' dell'Universit\`{a}
and Gruppo Collegato INFN, Salerno, Italy
\\
$^{31}$Dipartimento DISAT del Politecnico and Sezione INFN, Turin, Italy
\\
$^{32}$Dipartimento di Scienze e Innovazione Tecnologica dell'Universit\`{a} del Piemonte Orientale and INFN Sezione di Torino, Alessandria, Italy
\\
$^{33}$Dipartimento Interateneo di Fisica `M.~Merlin'
and Sezione INFN, Bari, Italy
\\
$^{34}$Division of Experimental High Energy Physics, University of Lund, Lund, Sweden
\\
$^{35}$European Organization for Nuclear Research (CERN), Geneva, Switzerland
\\
$^{36}$Excellence Cluster Universe, Technische Universit\"{a}t M\"{u}nchen, Munich, Germany
\\
$^{37}$Faculty of Engineering, Bergen University College, Bergen, Norway
\\
$^{38}$Faculty of Mathematics, Physics and Informatics, Comenius University, Bratislava, Slovakia
\\
$^{39}$Faculty of Nuclear Sciences and Physical Engineering, Czech Technical University in Prague, Prague, Czech Republic
\\
$^{40}$Faculty of Science, P.J.~\v{S}af\'{a}rik University, Ko\v{s}ice, Slovakia
\\
$^{41}$Faculty of Technology, Buskerud and Vestfold University College, Tonsberg, Norway
\\
$^{42}$Frankfurt Institute for Advanced Studies, Johann Wolfgang Goethe-Universit\"{a}t Frankfurt, Frankfurt, Germany
\\
$^{43}$Gangneung-Wonju National University, Gangneung, South Korea
\\
$^{44}$Gauhati University, Department of Physics, Guwahati, India
\\
$^{45}$Helmholtz-Institut f\"{u}r Strahlen- und Kernphysik, Rheinische Friedrich-Wilhelms-Universit\"{a}t Bonn, Bonn, Germany
\\
$^{46}$Helsinki Institute of Physics (HIP), Helsinki, Finland
\\
$^{47}$Hiroshima University, Hiroshima, Japan
\\
$^{48}$Indian Institute of Technology Bombay (IIT), Mumbai, India
\\
$^{49}$Indian Institute of Technology Indore, Indore, India
\\
$^{50}$Indonesian Institute of Sciences, Jakarta, Indonesia
\\
$^{51}$Inha University, Incheon, South Korea
\\
$^{52}$Institut de Physique Nucl\'eaire d'Orsay (IPNO), Universit\'e Paris-Sud, CNRS-IN2P3, Orsay, France
\\
$^{53}$Institute for Nuclear Research, Academy of Sciences, Moscow, Russia
\\
$^{54}$Institute for Subatomic Physics of Utrecht University, Utrecht, Netherlands
\\
$^{55}$Institute for Theoretical and Experimental Physics, Moscow, Russia
\\
$^{56}$Institute of Experimental Physics, Slovak Academy of Sciences, Ko\v{s}ice, Slovakia
\\
$^{57}$Institute of Physics, Academy of Sciences of the Czech Republic, Prague, Czech Republic
\\
$^{58}$Institute of Physics, Bhubaneswar, India
\\
$^{59}$Institute of Space Science (ISS), Bucharest, Romania
\\
$^{60}$Institut f\"{u}r Informatik, Johann Wolfgang Goethe-Universit\"{a}t Frankfurt, Frankfurt, Germany
\\
$^{61}$Institut f\"{u}r Kernphysik, Johann Wolfgang Goethe-Universit\"{a}t Frankfurt, Frankfurt, Germany
\\
$^{62}$Institut f\"{u}r Kernphysik, Westf\"{a}lische Wilhelms-Universit\"{a}t M\"{u}nster, M\"{u}nster, Germany
\\
$^{63}$Instituto de Ciencias Nucleares, Universidad Nacional Aut\'{o}noma de M\'{e}xico, Mexico City, Mexico
\\
$^{64}$Instituto de F\'{\i}sica, Universidad Nacional Aut\'{o}noma de M\'{e}xico, Mexico City, Mexico
\\
$^{65}$Institut Pluridisciplinaire Hubert Curien (IPHC), Universit\'{e} de Strasbourg, CNRS-IN2P3, Strasbourg, France
\\
$^{66}$iThemba LABS, National Research Foundation, Somerset West, South Africa
\\
$^{67}$Joint Institute for Nuclear Research (JINR), Dubna, Russia
\\
$^{68}$Konkuk University, Seoul, South Korea
\\
$^{69}$Korea Institute of Science and Technology Information, Daejeon, South Korea
\\
$^{70}$KTO Karatay University, Konya, Turkey
\\
$^{71}$Laboratoire de Physique Corpusculaire (LPC), Clermont Universit\'{e}, Universit\'{e} Blaise Pascal, CNRS--IN2P3, Clermont-Ferrand, France
\\
$^{72}$Laboratoire de Physique Subatomique et de Cosmologie, Universit\'{e} Grenoble-Alpes, CNRS-IN2P3, Grenoble, France
\\
$^{73}$Laboratori Nazionali di Frascati, INFN, Frascati, Italy
\\
$^{74}$Laboratori Nazionali di Legnaro, INFN, Legnaro, Italy
\\
$^{75}$Lawrence Berkeley National Laboratory, Berkeley, California, United States
\\
$^{76}$Moscow Engineering Physics Institute, Moscow, Russia
\\
$^{77}$Nagasaki Institute of Applied Science, Nagasaki, Japan
\\
$^{78}$National Centre for Nuclear Studies, Warsaw, Poland
\\
$^{79}$National Institute for Physics and Nuclear Engineering, Bucharest, Romania
\\
$^{80}$National Institute of Science Education and Research, Bhubaneswar, India
\\
$^{81}$National Research Centre Kurchatov Institute, Moscow, Russia
\\
$^{82}$Niels Bohr Institute, University of Copenhagen, Copenhagen, Denmark
\\
$^{83}$Nikhef, Nationaal instituut voor subatomaire fysica, Amsterdam, Netherlands
\\
$^{84}$Nuclear Physics Group, STFC Daresbury Laboratory, Daresbury, United Kingdom
\\
$^{85}$Nuclear Physics Institute, Academy of Sciences of the Czech Republic, \v{R}e\v{z} u Prahy, Czech Republic
\\
$^{86}$Oak Ridge National Laboratory, Oak Ridge, Tennessee, United States
\\
$^{87}$Petersburg Nuclear Physics Institute, Gatchina, Russia
\\
$^{88}$Physics Department, Creighton University, Omaha, Nebraska, United States
\\
$^{89}$Physics Department, Panjab University, Chandigarh, India
\\
$^{90}$Physics Department, University of Athens, Athens, Greece
\\
$^{91}$Physics Department, University of Cape Town, Cape Town, South Africa
\\
$^{92}$Physics Department, University of Jammu, Jammu, India
\\
$^{93}$Physics Department, University of Rajasthan, Jaipur, India
\\
$^{94}$Physikalisches Institut, Eberhard Karls Universit\"{a}t T\"{u}bingen, T\"{u}bingen, Germany
\\
$^{95}$Physikalisches Institut, Ruprecht-Karls-Universit\"{a}t Heidelberg, Heidelberg, Germany
\\
$^{96}$Physik Department, Technische Universit\"{a}t M\"{u}nchen, Munich, Germany
\\
$^{97}$Purdue University, West Lafayette, Indiana, United States
\\
$^{98}$Pusan National University, Pusan, South Korea
\\
$^{99}$Research Division and ExtreMe Matter Institute EMMI, GSI Helmholtzzentrum f\"ur Schwerionenforschung, Darmstadt, Germany
\\
$^{100}$Rudjer Bo\v{s}kovi\'{c} Institute, Zagreb, Croatia
\\
$^{101}$Russian Federal Nuclear Center (VNIIEF), Sarov, Russia
\\
$^{102}$Saha Institute of Nuclear Physics, Kolkata, India
\\
$^{103}$School of Physics and Astronomy, University of Birmingham, Birmingham, United Kingdom
\\
$^{104}$Secci\'{o}n F\'{\i}sica, Departamento de Ciencias, Pontificia Universidad Cat\'{o}lica del Per\'{u}, Lima, Peru
\\
$^{105}$Sezione INFN, Bari, Italy
\\
$^{106}$Sezione INFN, Bologna, Italy
\\
$^{107}$Sezione INFN, Cagliari, Italy
\\
$^{108}$Sezione INFN, Catania, Italy
\\
$^{109}$Sezione INFN, Padova, Italy
\\
$^{110}$Sezione INFN, Rome, Italy
\\
$^{111}$Sezione INFN, Trieste, Italy
\\
$^{112}$Sezione INFN, Turin, Italy
\\
$^{113}$SSC IHEP of NRC Kurchatov institute, Protvino, Russia
\\
$^{114}$Stefan Meyer Institut f\"{u}r Subatomare Physik (SMI), Vienna, Austria
\\
$^{115}$SUBATECH, Ecole des Mines de Nantes, Universit\'{e} de Nantes, CNRS-IN2P3, Nantes, France
\\
$^{116}$Suranaree University of Technology, Nakhon Ratchasima, Thailand
\\
$^{117}$Technical University of Ko\v{s}ice, Ko\v{s}ice, Slovakia
\\
$^{118}$Technical University of Split FESB, Split, Croatia
\\
$^{119}$The Henryk Niewodniczanski Institute of Nuclear Physics, Polish Academy of Sciences, Cracow, Poland
\\
$^{120}$The University of Texas at Austin, Physics Department, Austin, Texas, United States
\\
$^{121}$Universidad Aut\'{o}noma de Sinaloa, Culiac\'{a}n, Mexico
\\
$^{122}$Universidade de S\~{a}o Paulo (USP), S\~{a}o Paulo, Brazil
\\
$^{123}$Universidade Estadual de Campinas (UNICAMP), Campinas, Brazil
\\
$^{124}$Universidade Federal do ABC, Santo Andre, Brazil
\\
$^{125}$University of Houston, Houston, Texas, United States
\\
$^{126}$University of Jyv\"{a}skyl\"{a}, Jyv\"{a}skyl\"{a}, Finland
\\
$^{127}$University of Liverpool, Liverpool, United Kingdom
\\
$^{128}$University of Tennessee, Knoxville, Tennessee, United States
\\
$^{129}$University of the Witwatersrand, Johannesburg, South Africa
\\
$^{130}$University of Tokyo, Tokyo, Japan
\\
$^{131}$University of Tsukuba, Tsukuba, Japan
\\
$^{132}$University of Zagreb, Zagreb, Croatia
\\
$^{133}$Universit\'{e} de Lyon, Universit\'{e} Lyon 1, CNRS/IN2P3, IPN-Lyon, Villeurbanne, Lyon, France
\\
$^{134}$Universit\`{a} di Brescia, Brescia, Italy
\\
$^{135}$V.~Fock Institute for Physics, St. Petersburg State University, St. Petersburg, Russia
\\
$^{136}$Variable Energy Cyclotron Centre, Kolkata, India
\\
$^{137}$Warsaw University of Technology, Warsaw, Poland
\\
$^{138}$Wayne State University, Detroit, Michigan, United States
\\
$^{139}$Wigner Research Centre for Physics, Hungarian Academy of Sciences, Budapest, Hungary
\\
$^{140}$Yale University, New Haven, Connecticut, United States
\\
$^{141}$Yonsei University, Seoul, South Korea
\\
$^{142}$Zentrum f\"{u}r Technologietransfer und Telekommunikation (ZTT), Fachhochschule Worms, Worms, Germany
\endgroup

\end{document}